\documentclass[a4paper, twocolumn, twoside,
superscriptaddress,
nofootinbib,
amsmath,amssymb,
pra,
]{revtex4-2}
\usepackage{amsmath,amssymb, amsthm, amsfonts}
\usepackage{dsfont} 
\usepackage{xspace}
\usepackage{xcolor}
\usepackage{changes}
\usepackage{verbatim}
\usepackage{float}
\usepackage{comment}
\usepackage{soul}
\usepackage{multirow}
\usepackage{color}
\usepackage{graphicx}
\usepackage{subfig}
\usepackage[f]{esvect}
\usepackage{accents}
\usepackage{bm}
\usepackage{comment}
 \usepackage[unicode=true, breaklinks=false, pdfborder={0 0 1}, backref=false, colorlinks=true, linkcolor=blue, citecolor=blue, urlcolor=blue]{hyperref}

\DeclareMathAlphabet{\mathbbold}{U}{bbold}{m}{n}

\newcommand*{\rttensor}[1]{\bar{\bar{#1}}}

\newcommand{\ket}[1]{\vert #1\rangle}
\newcommand{\bra}[1]{\langle#1\vert}
\newcommand{\bb}[0]{\begin{eqnarray}}
\newcommand{\ee}[0]{\end{eqnarray}}

\begin{document}


\title{Decoherence induced by dipole-dipole couplings between atomic species in rare-earth ion-doped Y$_2$SiO$_5$.}

\author{C. Pignol}
\affiliation{Universit\'e C\^ote d'Azur, CNRS, Institut de Physique de Nice, 06200 Nice, France}
\author{A. Ortu}
\affiliation{Institut f\"ur Quantenoptik und Quanteninformation, \"Osterreichische Akademie der Wissenschaften, Technikerstrasse 21a, 6020 Innsbruck, Austria}
\affiliation{Universit\"at Innsbruck, Institut f\"ur Experimentalphysik, Technikerstrasse 25, 6020 Innsbruck, Austria}
\author{L. Nicolas}
\affiliation{Department of Applied Physics, University of Geneva, CH-1211 Geneva, Switzerland}
\author{V. D'Auria}
\affiliation{Universit\'e C\^ote d'Azur, CNRS, Institut de Physique de Nice, 06200 Nice, France}
\author{S. Tanzilli}
\affiliation{Universit\'e C\^ote d'Azur, CNRS, Institut de Physique de Nice, 06200 Nice, France}
\author{T. Chaneli\`ere}
\affiliation{Université Grenoble Alpes, CNRS, Grenoble INP, Institut N\'eel, 38000, Grenoble, France}
\author{M. Afzelius}
\affiliation{Department of Applied Physics, University of Geneva, CH-1211 Geneva, Switzerland}
\author{J. Etesse}
\affiliation{Universit\'e C\^ote d'Azur, CNRS, Institut de Physique de Nice, 06200 Nice, France}
\email{jean.etesse@univ-cotedazur.fr}
\date{\today}

\begin{abstract}
Rare-earth ion doped crystals are state-of-the-art platforms for processing quantum information, particularly thanks to their excellent optical and spin coherence properties at cryogenic temperatures. Experimental observations have shown that the application of a static magnetic bias field significantly improves the coherence times in the rare-earth ions ensemble, but only a few studies have focused on its the dependency as a function of both magnetic field direction and amplitude. This is especially true for magnetic field amplitudes under the mT, and for low magnetic dipole moment ions. In this paper, we investigate the relationship between the magnetic field parameters and the decoherence caused by magnetic dipole-dipole coupling with the nearest neighbors nuclear spins in the crystal. The primary non-Kramers rare-earth ions investigated here are europium and praseodymium, but we also extend our study to the ytterbium Kramers ion due to its low magnetic dipole in the mT range. We perform theoretical investigations and simulations of the energy structure and coherence time evolution and identify good correspondences between experimental and simulated spin echo data. This work allows us to pinpoint the most relevant decoherence mechanisms in the considered magnetic field regime, and to predict favorable magnetic configurations.

\end{abstract}

\maketitle

\section{Introduction}
Quantum networks offer exciting prospects such as unconditionally secure communication through quantum key distribution~\cite{RevModPhys.74.145,ekert_ultimate_2014}, enhanced computational power via distributed quantum computing~\cite{PhysRevA.59.4249,8910635}, and ultra-sensitive measurements with quantum sensors~\cite{komar_quantum_2014,PhysRevLett.109.070503}. Furthermore, they hold promise for fundamental scientific investigations, including testing the foundations of quantum theory~\cite{hensen_loophole-free_2015,PhysRevLett.119.010402}.
The foreseen challenges for creating operational quantum networks include (i) maintaining quantum states over distances above hundreds of kilometers despite propagation losses in fibers~\cite{Inagaki:13}, (ii) managing decoherence and noise~\cite{PhysRevX.6.021040}, and (iii) developing reliable protocols for quantum operations and information transfer~\cite{10.1145/3386367.3431293}.\par
This is where quantum memories set themselves as crucial elements. They can act as quantum repeaters to extend the distance over which quantum information can be propagated~\cite{PhysRevLett.81.5932,hermans_qubit_2022, Guo2022}, to delay the information transmission for protocols that require synchronization~\cite{PhysRevLett.118.220501}, as well as to enable single photon generation~\cite{PhysRevLett.103.113602} and perform operations on qubits~\cite{doi:10.1126/science.aah4758}. Many platforms have been envisioned to implement quantum memories, such as cold atoms~\cite{chaneliere_storage_2005}, hot atomic vapors~\cite{PRXQuantum.3.020349, PhysRevA.97.042316}, solid-state artificial atoms~\cite{PhysRevA.97.033823}, and all-optical storage loops~\cite{Bouillard2018QuantumSO}. Amongst all these possibilities, rare-earth (RE) ion doped crystals are recognized as state-of-the-art candidates. Moreover, such material also allow, among others, for photonic information processing~\cite{Craiciu:21,Kinos22}, single-spin manipulation for microwave processing~\cite{Wang_2023} as well as opto-microwave transduction~\cite{PhysRevA.99.063830}.\par Their principal interest lies in their intrinsically long optical and spin coherence times~\cite{Zhong2015}, allowing to perform storage of up to an hour~\cite{Ma2021}. Furthermore, their large optical inhomogeneous broadening offers great multiplexing capabilities which allow handling thousands of modes in the spectro-temporal domain~\cite{Businger2022}. One means to achieve long spin coherence time is to operate in a very specific magnetic regime called Zero First-Order Zeeman (ZEFOZ) point (also referred to as clock transitions), where the magnetic sensitivity to the environmental fluctuations is minimal~\cite{PhysRevLett.111.033601}. The downsides of working in this regime are that it usually requires to operate in the 100~mT-10~T amplitude range, and to adjust the field direction with very high precision~\cite{zhong_development_2017}. In contrast, recent experiments have shown that the application of a weak magnetic field ($<10~$mT) already suffices to strongly affect the spin coherence times of the ensembles~\cite{ortu_storage_2022,Holzäpfel_2020}. Depending on the RE species, such effect can lead either to an increase~\cite{Equall1994,ortu_storage_2022,Holzäpfel_2020} or to a decrease~\cite{Nicolas2022} of the spin coherence time. \par
Even if each RE ion ensemble behaves differently in different magnetic field regimes, we however postulate that the underlying mechanisms that lead to spin decoherence are the same for all species and stem from magnetic dipole-dipole coupling~\cite{Fraval2004b}. Several approaches have been used to explore this question, using either averaging techniques, perturbative treatments, and stochastic models, that enabled the models to be effective within a specific frame of study~\cite{ PhysRev.125.912,Mims1968,PhysRevB.79.115320,Ma2023}. More recently, a microscopic model based on dipole-dipole interaction in erbium allowed to replicate complex decay patterns~\cite{Car2020} and selective spin bath ion addressing~\cite{PhysRevB.100.165107}. This approach, directly linked to cluster correlation expansion (CCE) methods~\cite{Witzel06,PhysRevB.78.085315,PhysRevB.79.115320}, can however not be used with all ions in all magnetic field regimes. Inspired by this approach, we present here a general and predictive model that can be applied to any ion species in any host matrix, allowing one to estimate the effect of magnetic dipole-dipole interaction on coherence time.\par
The main novelty of our approach is that it does not rely on particular coupling or dynamic hypothesis but on an exact simulation of spin dynamics in interaction with a few ($\sim$~5) nearest neighbors nuclear spins.
To this end, we adopt a microscopic approach and theoretically investigate the impact of magnetic dipole-dipole couplings between the host matrix ions (that possess a nuclear spin) and the RE ion on the coherence time. In our study, we focus on three RE ions: europium $\mathrm{^{151}Eu^{3+}}$ (Eu), ytterbium $\mathrm{^{171}Yb^{3+}}$ (Yb) and praseodymium $\mathrm{^{141}Pr^{3+}}$ (Pr). Here, we consider the yttrium orthosilicate matrix ($\mathrm{Y_{2}SiO_{5}}$, hereafter noted YSO), in which the RE ions substitute Y$^{3+}$ ions. Such a crystal is commonly used due to its low nuclear spin density \cite{zhong_development_2017,Equall1994,PhysRevLett.92.077601}. We adopt three different approaches for the three RE at stake for confrontation with our numerical model. For europium, we perform an experiment based on the atomic frequency comb (AFC) protocol~\cite{AFC} to estimate the spin coherence time. For ytterbium, we use already published experimental data. Finally, for praseodymium, we only perform numerical simulations to predict a favorable magnetic configuration.
As stated above, the considered magnetic field amplitude ranges from the $\mu$T to the mT.  \par
This article is organized as follows. First, we present the theoretical model of the problem in Sec. \ref{model}, by introducing the model itself in Sec. \ref{ex_model}, and its possible alternatives and approximations in Sec. \ref{approx_sec}. We then present our simulations and findings in Sec. \ref{res_sec}. We compare our simulations with experimental results for Eu in Sec. \ref{Eu_sec}, then for Yb in Sec. \ref{Yb_sec}. We continue by performing coherence time predictions for the Pr ion, as a function of magnetic field amplitude and orientation in Sec. \ref{Pr_sec}. In Sec. \ref{Y_sec}, we summarize and explicit the relationship between the RE-Y dipole-dipole interaction, the magnetic field and the coherence time. Finally, we present in Sec. \ref{persp_sec} the directions for future investigations.  

\section{Model}\label{model}
\subsection{Simulated system and Hamiltonian}\label{ex_model}

\begin{figure*}
  \includegraphics[width=1.8\columnwidth]{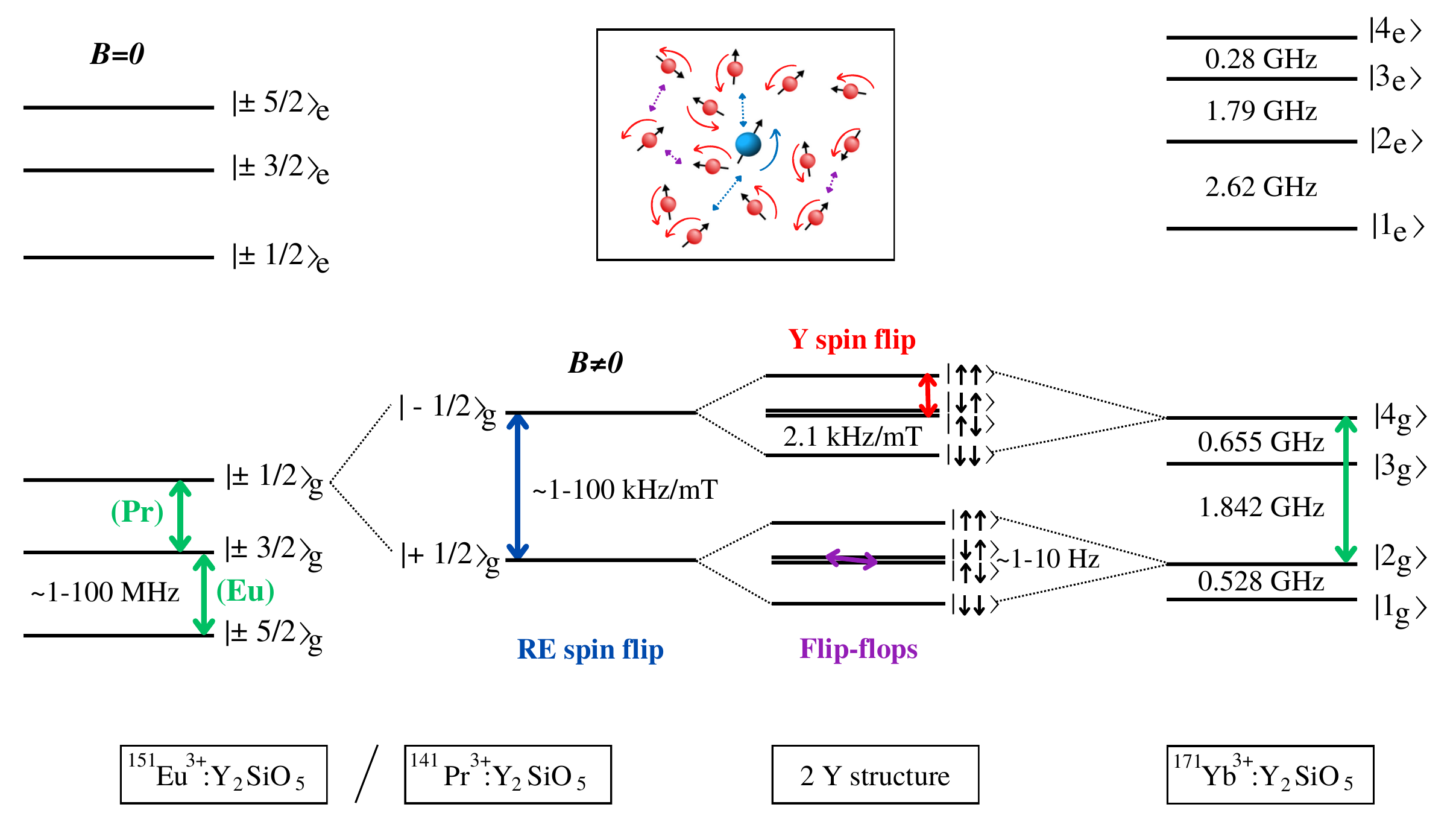}
\caption{Schematic energy levels of Eu:YSO, Pr:YSO and Yb:YSO with the energetic structure of the spin nuclear bath $\hat{H}_{\rm Y}$ for two Y ions. The green arrows highlight the spin transitions under study. Blue, red, and purple arrows in the top inset indicate spin dynamics mechanisms discussed in the body of the text.}
\label{fig:data1}
\end{figure*}

We show in this article that the decoherence mechanisms in the RE ion-doped crystal under study can be investigated by focusing on a Hilbert space of a rather small dimension restricted to the central RE ion plus a few nearest neighbors nuclear spins in the bath. For that, the concentration in doping RE ions has to be low enough to be able to neglect the magnetic dipole-dipole coupling between themselves. Because of their strong electronic moment, Kramers ions need a much lower concentration to satisfy this condition than non-Kramers ions. We can cite as an example erbium-doped crystals at 10~ppm \cite{Car2018}, meanwhile a 1000~ppm concentration is viable for an europium-doped crystal \cite{PhysRevA.103.022618}.\par
Without interactions amongst the RE ions, the crystal is then considered as constituted of randomly distributed independent and identical cells containing one RE ion surrounded by yttrium (Y) ions. We simulate one of these cells, but because of computation power limitations we restrict the number of Y ions inside to a maximum of seven Y ions. The Y ions are added to the system from the closest to the RE ion to the furthest, and their positions are taken from published computations made from the crystal lattice parameters \cite{Car2020}.\par
YSO has a monoclinic structure and belongs to the $C_{2h}^{6}$ space group. Crystals are usually cut along the $({D_{1}}, {D_{2}},{b})$ axes that correspond to the polarization extinction axes \cite{135248}. The RE ions can substitute the Y ions in two different optical sites, I and II \cite{KURKIN1980233}. Each optical site consists of two magnetic sub-sites that are related to each other via the $C_{2}$ symmetry axis (which coincides with the ${b}$ axis). For a magnetic field direction in the $({D_{1}}, {D_{2}})$ plane or along $b$, these magnetic sub-sites become equivalent. Most of the simulations presented in the following involve fields in this plane.\par

\subsubsection{The rare-earth ion}
The RE ion Hamiltonians of non-Kramers (such as Eu and Pr) and Kramers ions (such as Yb) are different. They lead to the energetic structures displayed in Fig.~\ref{fig:data1}. Eu- and Pr- doped YSO have an Hamiltonian of the form \cite{PhysRevB.97.094416,PhysRevLett.92.077601}:
\begin{equation}
\hat{H}_{\rm nK}= \mathbf{\hat{I}}\cdot\rttensor{Q}\cdot\mathbf{\hat{I}} + \mathbf{B}_{\rm DC}\cdot\rttensor{M}\cdot\mathbf{\hat{I}} + \mathbf{B}_{\rm DC}\cdot\rttensor{Z}\cdot\mathbf{B}_{\rm DC},
\label{eq:nonKramersH0}
\end{equation}
with the static magnetic field vector $\mathbf{B}_{\rm DC}$ expressed in the $({D_{1}}, {D_{2}},{b})$ basis as:
\begin{equation}\label{B}
\mathbf{B}_{\rm DC}=B_{\rm DC} \begin{pmatrix}\cos(\varphi)\sin(\theta) \\ \sin(\varphi)\sin(\theta) \\ \cos(\theta) \end{pmatrix}.
\end{equation}
In Eq.~(\ref{eq:nonKramersH0}), $\mathbf{\hat{I}}$ is the nuclear spin vector, $\rttensor{Q}$ the quadrupolar tensor, $\rttensor{M}$ the linear Zeeman tensor and $\rttensor{Z}$ the quadratic Zeeman tensor. The quadrupolar interaction is present for spins $I \geqslant 1$, and both Pr and Eu ions have an $I=5/2$ spin. The Zeeman interaction lifts the degeneracy of the quadrupolar levels by an amount dependent on the magnetic field orientation and amplitude. The order of magnitude of this effect is $\sim1$ to $10$~kHz/mT for Eu and $\sim10$ to $100$~kHz/mT for Pr, as illustrated in Fig.~\ref{fig:data1}. Then, the quadratic Zeeman tensor has (for instance with Eu) a magnitude on the order of $100~$Hz/mT$^{2}$ \cite{zhong_development_2017}. This means that the induced dipole is below $100~$Hz/mT in the magnetic field amplitude range considered here, far smaller than the $\sim1$ to $10~$kHz/mT induced by $\rttensor{M}$. Consequently, in the investigated magnetic field range ($\mu T$-mT), we can reasonably neglect the quadratic Zeeman interaction term. 
The energy structure created by the $\hat{H}_{\rm nK}$ Hamiltonian is schematically displayed in Fig.~\ref{fig:data1} left. The optical electronic transition addressed is a $4f \longleftrightarrow 4f$ transition ($^{3}H_{4} \longleftrightarrow {}^{1}D_{2}$ for Pr at 605.98~nm and $^{5}D_{0} \longleftrightarrow {}^{7}F_{0}$ for Eu at 580.04~nm), for the crystallographic site I of both systems. The spin transitions under study are the radiofrequency (RF) $\ket{\pm3/2}_g\leftrightarrow\ket{\pm1/2}_g$ 10.2~MHz transition for Pr and the $\ket{\pm5/2}_g\leftrightarrow\ket{\pm3/2}_g$ 46.2~MHz transition for Eu (green arrows in Fig.~\ref{fig:data1} left). \par
The Yb Kramers ion Hamiltonian is \cite{PhysRevB.98.195110}:
\begin{equation}
\hat{H}_{\rm K}= \mathbf{\hat{I}}\cdot\rttensor{\mathcal{A}}\cdot\mathbf{\hat{S}}+ \mu_{B}\mathbf{B}_{\rm DC}\cdot\rttensor{g}\cdot\mathbf{\hat{S}} + \mu_{n}\cdot g_{n}\cdot\mathbf{B}_{\rm DC}\cdot\mathbf{\hat{I}}.
\end{equation}
The Yb electronic $1/2$ spin $\mathbf{\hat{S}}$ is coupled to its nuclear $1/2$ spin $\mathbf{\hat{I}}$ through the hyperfine interaction tensor $\rttensor{\mathcal{A}}$. It creates a set of four non-degenerate levels when doping YSO, that are separated by a frequency splitting wide enough to selectively address one transition optically. $\mu_{B}$ and $\mu_{n}$ are the Bohr magneton and the nuclear magneton, respectively, while $\rttensor{g}$ and $g_{n}$ are the electronic Zeeman interaction tensor and nuclear Zeeman interaction coefficient, respectively.\par
The energetic structure of Yb:YSO is schematically displayed on the right of Fig.~\ref{fig:data1}. The electronic optical transition for Yb is the ${}^{2}F_{7/2}(0) \longleftrightarrow {}^{2}F_{5/2}(0)$ at 978.85~nm for the crystallographic site II. The spin transition that we consider here is the $\ket{2_{g}} \longleftrightarrow \ket{4_{g}}$ microwave (MW) transition at $2.497$~GHz (green arrow on the right of Fig.~\ref{fig:data1}), whose spin decoherence was investigated recently~\cite{Nicolas2022}.\par

\subsubsection{The bath}

The RE ion is considered to be surrounded by $N$ yttrium ions, sole ions of the YSO matrix that possess a nuclear spin with non-negligible abundance (we neglect the contribution of $^{29}$Si, only present at 4.7\%~\cite{Equall1995,Konz2003}). The nuclear spin bath has an Hamiltonian of the form:
\begin{align}
\hat{H}_{\rm Y}=\sum_{k=1}^N\hat{H}_{\rm Y_k}^{\rm Z}+\sum_{i\neq j=1}^N\hat{H}_{\rm Y_i-Y_j}^{\rm dd}.
\label{bath_Hamilton}
\end{align}
The first term in this expression is the sum of the Zeeman Hamiltonians of each Y ion labeled $k$:
\begin{align}
\hat{H}_{\rm Y_k}^{\rm Z}=\gamma_{\rm Y}\mathbf{B}_{\rm DC}\cdot\mathbf{\hat{I}}_{1/2}^k,
\end{align}
where $\gamma_{\rm Y}=2$~kH/mT is the magnetic dipole moment of yttrium and $\mathbf{\hat{I}}_{1/2}^k$ is the 1/2 spin vector operator.\\
The second term $\hat{H}_{\rm Y_i-Y_j}^{\rm dd}$ is the dipole-dipole interaction of all the Y ions with each other. The general expression of such an Hamiltonian for two ions $1$ and $2$ whose positions are linked with the vector $\mathbf{r}_{12}$ is~\cite{Abragam1961}:
\begin{equation}
\hat{H}^{\rm dd}_{1-2}=-\frac{\mu_{0}h}{4\pi r_{12}^{3}}(3(\mathbf{\hat{m}}_{1}\cdot\mathbf{r}_{12})\otimes(\mathbf{\hat{m}}_{2}\cdot\mathbf{r}_{12}) - \mathbf{\hat{m}}_{1}\otimes\mathbf{\hat{m}}_{2}),
\label{dip-dip_gen}
\end{equation}
where $\mathbf{\hat{m}}_{1}$ and $\mathbf{\hat{m}}_{2}$ are the magnetic moments of the ions, $h$ the Planck constant, $\mu_{0}$ the vacuum magnetic permeability and $r_{12}=||\mathbf{r}_{12}||$. The Y-Y dipole-dipole interaction lifts the degeneracy of each Y Zeeman level. As they depend on the relative distance and orientation of each pair of ions in the crystalline matrix, they create a manifold of energy levels, whose energy separation is on the order of a few Hz that is different for each level pair. The energetic structure associated with this Hamiltonian is displayed in the center of Fig.~\ref{fig:data1}, for two Y ions. Total spin projection configurations have different energies due to the dipole-dipole coupling, and this ultimately leads to an apparent energetic broadening of the central spin transition, as will be illustrated later. Then, we define
\begin{align}
\hat{H}_0=\hat{H}_{\rm RE}+\hat{H}_{\rm Y}
\end{align}
the diagonalized Hamiltonian defining the energetic structure on which the dynamical evolution will next be considered.
We chose to include the Y-Y dipole-dipole interaction in this diagonalized Hamiltonian, leading to a lift of degeneracy in the bath manifolds as illustrated in Fig~\ref{fig:data1}. Such a choice is motivated by the fact that we want to focus on the effect of the bath as a whole on the RE ion through the RE-Y dipole-dipole couplings as we will introduce in the next paragraph. Notice that choosing to include Y-Y dipole-dipole interaction in the diagonalized part does not mean that we only consider diagonal interaction terms, as sometimes done in the literature~\cite{Herzog1956}.\\

\subsubsection{Dynamics}

Now that we have defined the energetic structure of the system, we focus on its dynamics. To this extent, we add a non-diagonal interaction Hamiltonian $\hat{H}_{\rm int}$ written in the eigenbasis of $\hat{H}_0$ to define the system total Hamiltonian:
\begin{equation}
\hat{H}=\hat{H}_{0}+\hat{H}_{\rm int}.
\label{eq:tot_hamilton}
\end{equation}
This interaction Hamiltonian consists of two contributions:
\begin{align}
\hat{H}_{\rm int}=\hat{H}^{\rm RF}_{\rm RE}+\sum_{k=1}^N\hat{H}_{\rm RE-Y_k}^{\rm dd}.
\label{eq:interaction_gal}
\end{align}
The first term $\hat{H}^{\rm RF}$ represents the resonant radiofrequency or microwave coupling via Zeeman interaction, which reads
\begin{align}
\hat{H}^{\rm RF}_{\rm nK}=\mathbf{B}_{\rm AC}\cdot\rttensor{M}\cdot\mathbf{\hat{I}}_{\rm RE}
\end{align}
for Eu and Pr, and
\begin{align}
\hat{H}^{\rm RF}_{\rm K}=\mu_{B}\mathbf{B}_{\rm AC}\cdot\rttensor{g}\cdot\mathbf{\hat{S}}
\end{align}
for Yb. These Hamiltonians drive the spin transitions (green arrows in Fig.~\ref{fig:data1}) during the echo sequences and the spin rephasing protocols, via the application of an oscillating magnetic field $\mathbf{B}_{\rm AC}$ resonant with the spin transition.\\
The second interaction term $\sum\hat{H}_{\rm RE-Y_k}^{\rm dd}$ describes the interaction of the central RE ion with the Y ions labeled by $k$. This Hamiltonian is also of the form (\ref{dip-dip_gen}), but now with different species in interaction. This contribution is the one that leads to decoherence in the sequences considered here.

\subsection{Framework for the simulations}\label{approx_sec}

In this section, we precisely define how decoherence is modeled, and discuss approximations that are to be taken in the previously defined Hamiltonians. We also compare them with different approximations usually considered in the literature for similar treatment, and explicit why most of them cannot be taken here.

\subsubsection{Approximations in our model}
\label{subsec_approx}
In order to understand which approximations can be made on our system, it is instructive to first rewrite the general dipole-dipole interaction Hamiltonian (\ref{dip-dip_gen}) as~\cite{Abragam1961}:
\begin{equation}\label{secularapprox}
\hat{H}^{\rm dd}_{1-2}= -\frac{\mu_{0}h}{4\pi r_{12}^{3}}(\hat{A}+\hat{B}+\hat{C}+\hat{C}^{\dagger}+\hat{D}+\hat{D}^{\dagger}),
\end{equation}
with:
\begin{align*}
\hat{A} =& (1-3\cos^{2}(\theta_{12})){\hat{m}}_{1}^z{\hat{m}}_{2}^z,\\
\hat{B} =& -\frac{1}{4}(1-3\cos^{2}(\theta_{12}))({\hat{m}}_{1}^{+}{\hat{m}}_{2}^{-}+{\hat{m}}^{-}_{1}{\hat{m}}_{2}^{+}),\\
\hat{C} =& -\frac{3}{2}\sin(\theta_{12})\cos(\theta_{12})e^{-i\varphi_{12}}({\hat{m}}^{+}_{1}{\hat{m}}_{2}^z+{\hat{m}}_{1}^z{\hat{m}}^{+}_{2}),\\
\hat{D} =& -\frac{3}{4}\sin^{2}(\theta_{12})e^{-2i\varphi_{12}}{\hat{m}}^{+}_{1}{\hat{m}}^{+}_{2}.
\end{align*}
In this expression, $\theta_{12}$ and $\varphi_{12}$ are the polar angles of the $\mathbf{r_{12}}$ vector that connects ions $1$ and $2$ with respect to the $z$-axis and ${\hat{m}}_{k}^{\pm}=\hat{m}_{k}^x\pm i\hat{m}_{k}^y$ the ladder operator of spin $k$. \par The first term $\hat{A}$ adds to each energy level of the system a different frequency offset. This term is sometimes stamped as `z-coupling' and can induce single ion inhomogeneous broadening of transitions \cite{Mims1968}. The $\hat{B}$ term describes the mechanism through which the two interacting ions exchange their spin states, often referred to as flip-flop mechanism. For two identical ions, this mechanism is energy conserving, however it is not conserving for the RE-Y interaction. The $\hat{C}$ part allows one ion to change its spin state while the other remains unchanged.
Finally, the $\hat{D}$ term is the double spin flip mechanism.  Note that the conjugate interaction terms are included in $\hat{C}^{\dagger}$ and $\hat{D}^{\dagger}$. In our case, some of the mechanisms grasped with this picture, sketched in the inset of Fig.~\ref{fig:data1}, and highlighted with blue, red, and purple arrows, consist in:
\begin{enumerate}
\item[-] \textbf{Blue arrows: the central ion can spin-flip}, due to either flip-flop with one or several ions in the bath ($\hat{B}$ term in $\hat{H}^{\rm dd}_{\rm RE-Y_k}$, dashed blue arrow) or to an off-resonant single spin-flip ($\hat{m}_{\rm{Y_k}}^z\hat{m}^{\pm}_{\rm RE}$ terms in $\hat{H}^{\rm dd}_{\rm RE-Y_k}$, solid blue arrow);
\item[-] \textbf{Red arrows: the bath ions can spin-flip}, due to either off-resonant central ion coupling ($\hat{m}_{\rm{RE}}^z\hat{m}^{\pm}_{Y_k}$ terms in $\hat{H}^{\rm dd}_{\rm RE-Y_k}$) or to off-resonant bath coupling ($\hat{C}$ and $\hat{C}^\dagger$ terms in $\hat{H}^{\rm dd}_{\rm Y_i-Y_j}$);
\item[-] \textbf{Purple arrows: the bath ions can flip-flop}, due to resonant bath coupling ($\hat{B}$ term in $\hat{H}^{\rm dd}_{\rm Y_i-Y_j}$) or to central ion-mediated interaction ($\hat{m}_{\rm{RE}}^z\hat{m}^{+}_{\rm Y_i}$ with $\hat{m}_{\rm{RE}}^z\hat{m}^{-}_{\rm Y_j}$ involved respectively in $\hat{H}^{\rm dd}_{\rm RE-Y_i}$ and $\hat{H}^{\rm dd}_{\rm RE-Y_j}$).
\end{enumerate}
\begin{figure}[ht!]
\includegraphics[width=\columnwidth]{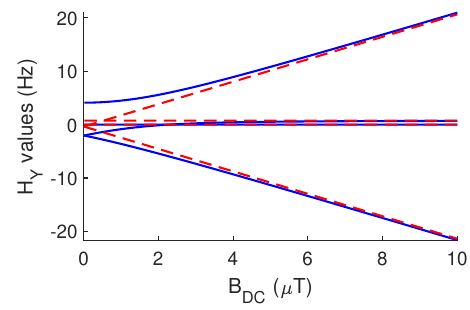}%
\caption{Eigenenergies of the bath $\hat{H}_{\rm Y}$ (given by Eq.(\ref{bath_Hamilton})) in the case $N=2$, as a function of the magnetic field magnitude.}
\label{fig:approx_ddY}
\end{figure}
In order to assess the strength of the different contributions in the Y bath interaction Hamiltonian $\hat{H}^{\rm dd}_{\rm Y_i-Y_j}$, we plot in Fig.~\ref{fig:approx_ddY} the eigenvalues of $\hat{H}_{\rm Y}$ for the first two Y neighbor ions in blue, and by removing the non-resonant contributions $\hat{C}$ and $\hat{D}$ in dashed red. We see that as soon as the ambient magnetic field $B_{\rm DC}$ is larger than a few micro-Tesla, the Zeeman splitting is already sufficient to prevent non-resonant processes to occur via direct dipole-dipole coupling. Therefore, in the following, we neglect these contributions and restrict the study only to z-coupling ($\hat{A}$ term) and flip-flop ($\hat{B}$ term) processes within the bath alone: this is the only hypothesis that we make in the following.\\
In the next section, we relate our work to considerations found in the literature, where hypothesis were taken to conduct estimations of coherence times, and show why we cannot map them to our study.

\subsubsection{Usual approximations not made here}
If we focus on the central ion, an usual and efficient way of estimating decay rates in the case of a system coupled to a quasi-continuum bath is to rely on the Fermi golden rule \cite{Syed22, PhysRevB.100.165107}. This has been for instance used to estimate flip-flop rates of Y spins in the bath \cite{Gong2017}. This approximation is frequent in models that follow a stochastic approach. The Y ions influence is often modeled as a mean fluctuating magnetic field emerging from the Y-Y dynamics \cite{Gong2017,Mims1968}, as a sudden jump model of Y spin changes \cite{PhysRev.125.912}, or as random noise \cite{Yang_2017}. All three models act as a perturbation that drives the decoherence on the central spin. Here we aim at estimating something slightly different: the central ion is coupled in a discrete manner to each state of the bath, with highly non-constant couplings between all levels. Moreover, no quasi-continuum hypothesis can be done in our case due to the purely deterministic landscape of the surrounding Y spins for each RE ion. Discrepancies between the Fermi golden rule and our model are addressed in App.~\ref{Appendix:Fermi}.\\
Another very common hypothesis done on the systems is the frozen central spin, in which spin flips of the central ion are neglected (mechanisms depicted with a blue arrow in Fig.~\ref{fig:data1}). In this case, the decoherence is estimated through the sole bath dynamics during an echo sequence. Interestingly, in this regime, explicit dynamics formula can be given by relying on methods such as the CCE \cite{Witzel06,PhysRevB.78.085315,PhysRevB.79.115320}. This technique relies on the decomposition of the bath into clusters of increasing size, allowing to infer the explicit dynamics within each cluster and to estimate with more and more precision the bath effect on the central spin when increasing the cluster size (and therefore the expansion order). For the simplest \textit{Loschmidt} echo decay, the decoherence over time is given by the overlap of the bath wavefunction with itself after free evolution whether the central spin is in its ground or excited state \cite{Rossini08}. In the present study, this approach cannot grasp the full dynamics of the system due to the low nuclear dipole moments of the central ions considered here. Indeed, the scaling difference between the Y nuclear moment ($\sim2$~kHz/mT) and Eu and Pr ($\sim10-100$~kHz/mT) is not large enough to prevent multi-spin contributions for flipping the central ion. In the low field regime, we will show that this is indeed the case. For Yb, the zero-field splitting however prevents these events and central ion flips can be neglected at low fields.\\
Recently, a mean-dipole model has also been developed to describe the echo decay dynamics in erbium-doped YSO \cite{Car2018,Car2020}. Here, the frozen central spin hypothesis is also made, and no bath interaction is considered. We show in App.~\ref{annex:frozen} that this approach is exactly a first-order CCE.\\
\subsubsection{The sequence under consideration}
\begin{figure}
\includegraphics[width=0.8\columnwidth]{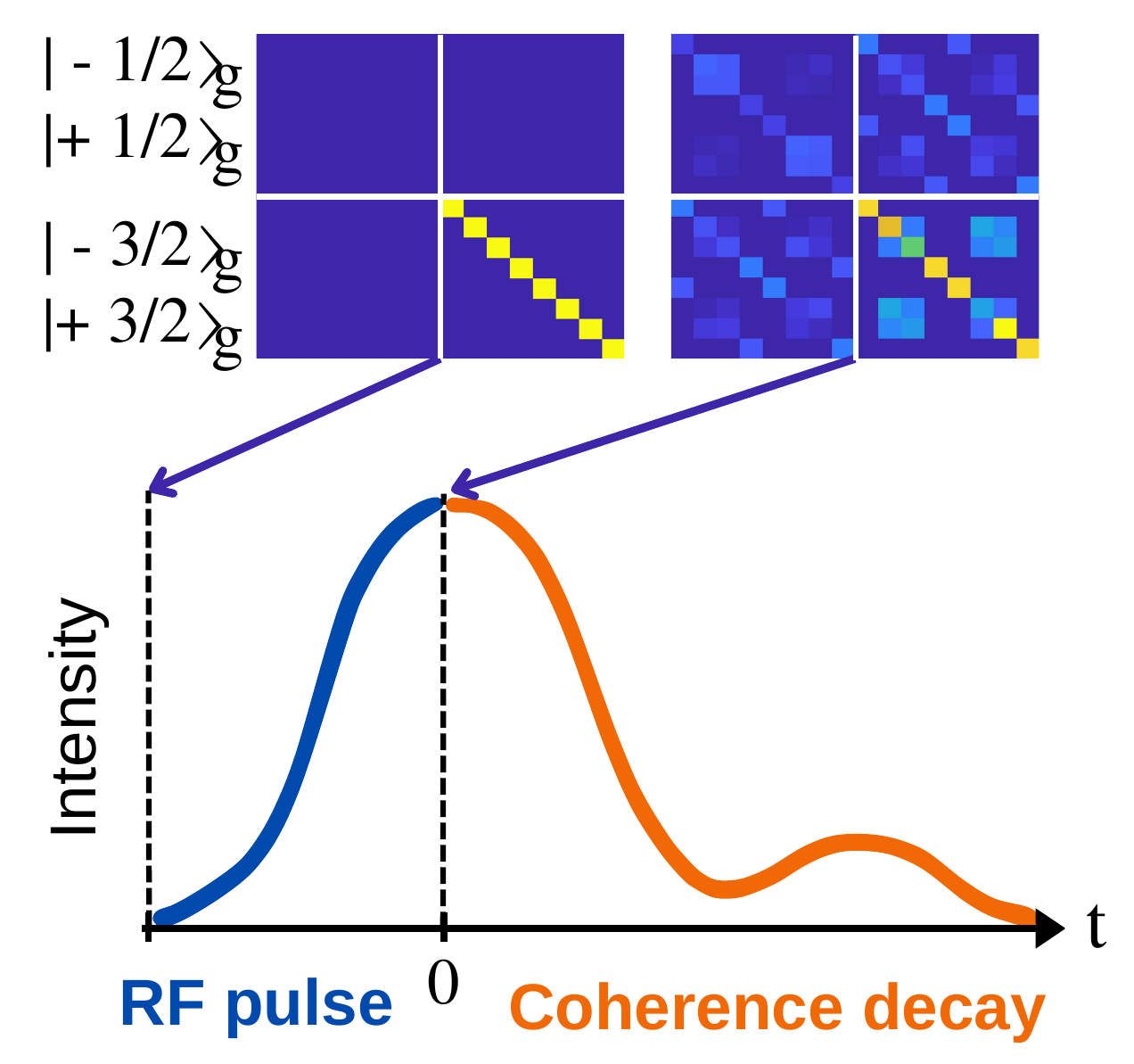}%
\caption{Sequence under study for simulating single-ion coherence decays: after creation of an initial coherence with a radiofrequency pulse, the coherence (\ref{echo}) is plotted as a function of time (coherence decay).}
\label{fig:time_seq}
\end{figure}
Finally, we define the sequence that we will consider to assess a coherence time. Our approach aims at modeling decoherence via the interaction of the central ion with a bath. As it is a single-spin decoherence process, no inhomogeity between RE ions of the ensembles is considered. Therefore, a simple free-induction decay (FID)-type sequence is used to simulate the effective decoherence of the central RE spin, in the same philosophy as when considering the previously mentioned Loschmidt echo. We however keep in mind that single-ion static broadening can be induced by sole z-coupling due to the $\hat{A}$ term in Eq.~(\ref{secularapprox}). This term effectively leads to frequency shifts of spin-preserving transitions that beat due to balanced statistical mixture of the initial bath population. To ensure that such effect is not limiting the coherence time, we can plot Hahn echo sequence and witness that similar coherence time is obtained (see App.~\ref{app:Hahn}). However, computation resources required for simulating such a sequence is very demanding, therefore we focus on FID sequences whenever possible.\\
The considered sequence is depicted in Fig.~\ref{fig:time_seq}. Initially, only the ground hyperfine level is populated and Zeeman (for Eu and Pr) and superhyperfine energy levels are in a balanced mixed state. A coherence is then created in the system with an RF or MW pulse on the relevant spin transition of the central RE ion (green arrows in Fig.~\ref{fig:data1}). This is illustrated in Fig.~\ref{fig:time_seq} with the representation of the absolute value of the density matrix before and after the RF pulse for one Pr ion and two Y bath ions. Note that in the subsequent plots we discard this time interval, such that we only plot the following decay, that we will from now on call the \textit{coherence decay} (time after $t=0$ in Fig.~\ref{fig:time_seq}).
Next, the system is left to evolve under the sole drive of the dipole-dipole interaction term of $\hat{H}_{\rm int}$. The evolution of the state vector can then be explicitly written at each instant:
\begin{align}
\ket{\psi(t)}&=\sum_n\sum_{k,l}\alpha_k(0)p_{n,l}p_{kl}^{*}e^{i\lambda_{l}t}\ket{n}\label{psit_freq}\\
&=\sum_n\alpha_n(t)\ket{n},
\end{align}
where the $\alpha_n(t)$ are the time-dependent components of the state vector of the system along the eigenvector $\ket{n}$ of $\hat{H}_0$. The elements $p_{i,j}$ and $\lambda_l$ are respectively the coefficients of the base change matrix $P$ and of the diagonal matrix $D$, such that, in its matrix form, the total Hamiltonian reads:
\begin{equation}\label{eigenbasis}
{H}=PDP^{-1},
\end{equation}
where the interaction Hamiltonian $\hat{H}_{\rm int}$ only consists in the dipole-dipole interaction (no RF drive during the FID).

The coherence decay intensity is then read optically at any time $t$ and computed as:
\begin{equation}\label{echo}
I(t)= \left|\rm{Tr} \left[ \hat{\rho}(t) \hat{\mu}^{\rm opt}\right]\right|^{2},
\end{equation}
where $\hat{\mu}^{\rm opt}$ is the optical moment of the RE ion. This formula emulates an optical readout of the coherence, as it is usually performed experimentally via the Raman heterodyne scattering (RHS) protocol~\cite{PhysRevLett.50.993}. Note that here we assume that the hyperfine population is maintained by the transfer pulse for optical readout (the density matrix in the spin transition is assumed to be the same as in the optical one). We however estimate that such consideration is valid for reproducing decay curves in qualitative agreement with experimental data.\\
Injecting expression (\ref{psit_freq}) in (\ref{echo}) gives the coherence decay in the pure state case:
\begin{align}
I(t)&=|\bra{\psi(t)}\hat{\mu}^{\rm opt}\ket{\psi(t)}|^2\nonumber\\
&=\left|\sum_{\substack{n,m\\k,l,k',l'}}\mu^{\rm opt}_{n,m}\alpha_k^*(0)\alpha_{k'}(0)p_{n,l}^*p_{k,l}p_{m,l'}p^*_{k',l'}e^{i(\lambda_{l'}-\lambda_{l})t}\right|^2\nonumber\\
&=\left|\sum_{l,l'}g_{l,l'}e^{i(\lambda_{l'}-\lambda_{l})t}\right|^2.
\label{analyt_dyn}
\end{align}
This expression shows that the frequencies appearing in the dynamics of the coherence decay are the interval between the eigenvalues of $\hat{H}$ noted $\Delta\lambda_{l,l'}=|\lambda_l-\lambda_{l'}|$ weighted by a function related to the coupling between the eigenvectors (coefficients $p_{i,j}$) and to the initial coherences $\alpha_k^*(0)\alpha_{k'}(0)$. We will use this formula to visualize relevant weighted frequencies appearing in the dynamics of the echo. From now on, frequencies $\Delta\lambda_{l,l'}$ are called the \textit{dynamical frequencies}.\\
Finally, when a coherence time shall be infered and displayed for a given coherence decay trace, we follow the procedure detailed in App.~\ref{App_A}. Such a procedure is required due to the very changing nature of the decay temporal profile with respect to magnetic field properties.\\

For the following sections, we perform our simulations on the three RE ions presented above. By identifying the dynamical frequencies at stake, we will for each case determine the predominant decoherence mechanism among the ones presented in Fig.~\ref{fig:data1}.

\section{Results}\label{res_sec}
\subsection{Europium}\label{Eu_sec}

\begin{figure}[h]
  \includegraphics[width=\linewidth]{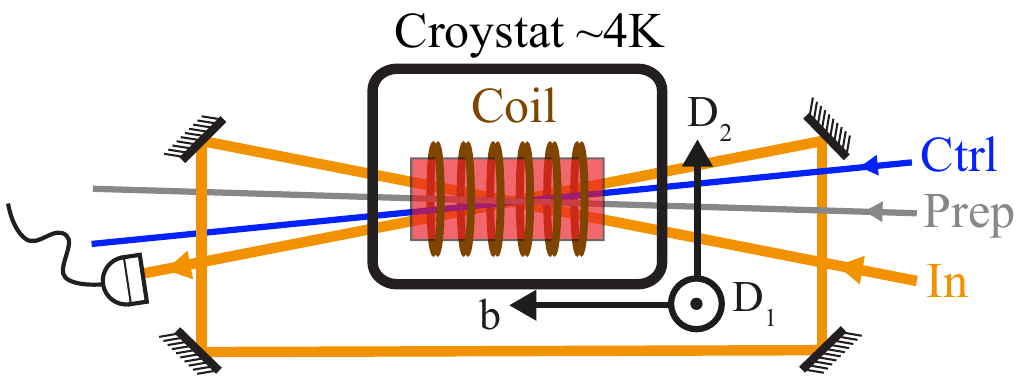}
  \caption{Experimental setup for the AFC spin-wave storage in Eu:YSO. See body of the text for details.}
  \label{fig:Euexp_setup}
\end{figure}

The first system under study is europium. We confront in the following our numerical simulations with experimentally acquired data points with the spin-wave AFC protocol, with a setup showed in Fig.~\ref{fig:Euexp_setup}. 

\begin{figure*}
\subfloat[ \label{fig:Eu_echoes}]{%
  \includegraphics[width=1\columnwidth]{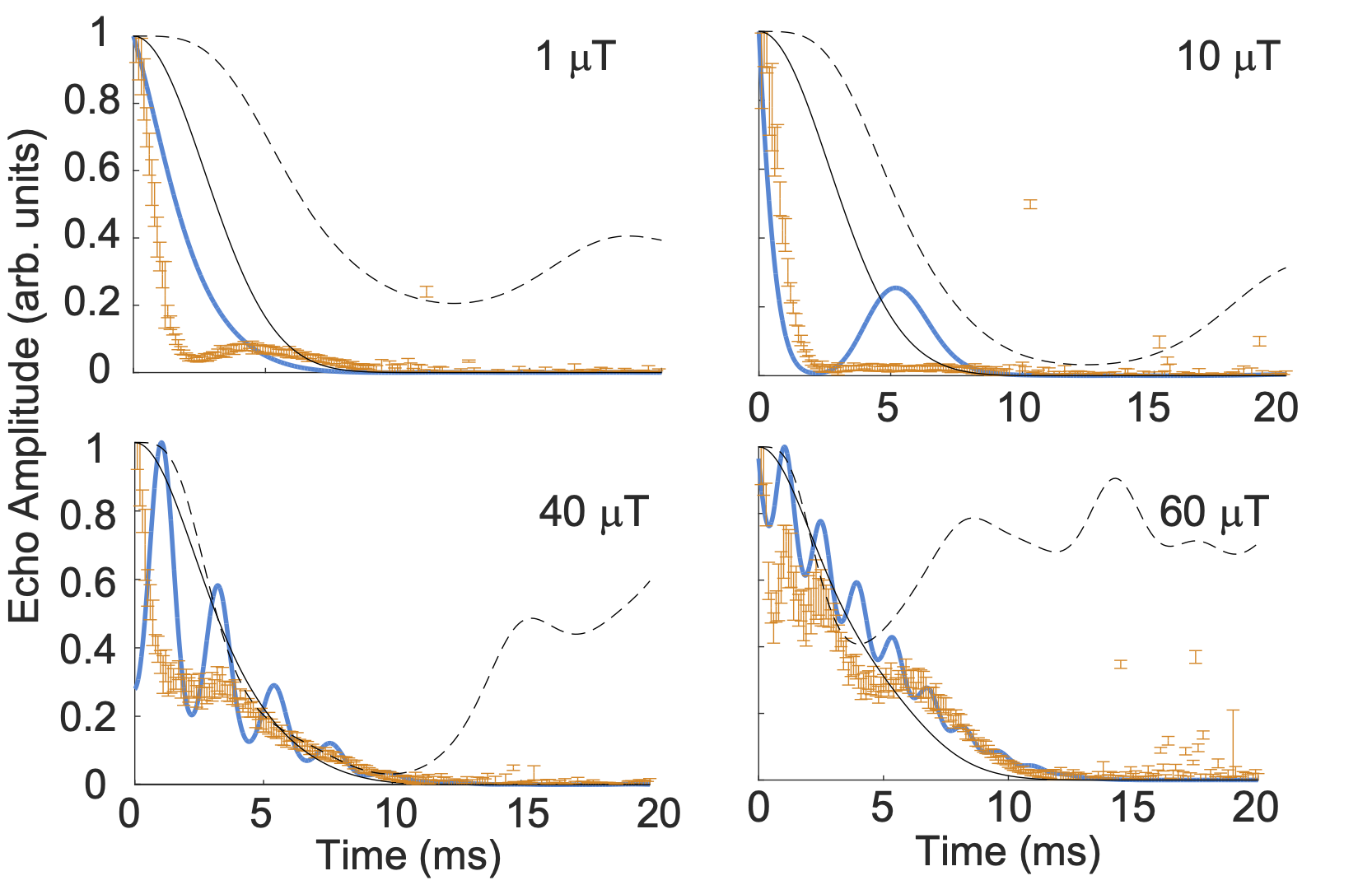}%
}\hfill
\subfloat[ \label{fig:2Y_Eu_freq_weights}]{%
  \includegraphics[width=1\columnwidth]{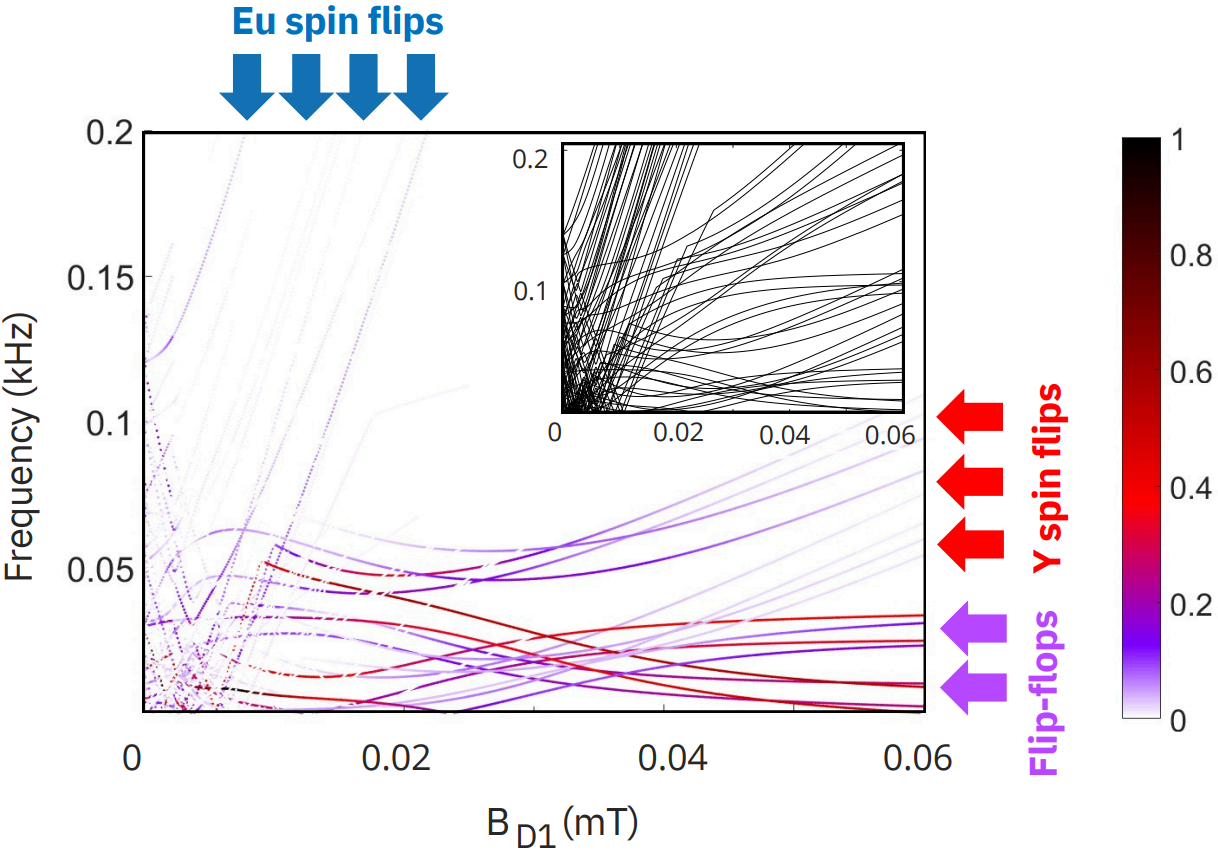}%
}\hfill
\caption{a) Experimental (orange dots) and simulated (blue lines) echo decay curves for Eu:YSO for the first $N=5$ neighboring Y ions, under a magnetic field oriented along $D_{1}$ set to an amplitude of $B_{\rm DC}=$ 1, 10, 40 and 60$~\mu$T. For comparison, decays obtained with first order CCE are shown in dash black (Hahn echo) and solid black (Loschmidt echo).  b) Simulation of the dynamical frequencies involved in the echo decays as a function of the magnetic field amplitude for Eu surrounded by two Y ions. The color scale represents the normalized $g$ weights (see Eq.~(\ref{analyt_dyn})) associated with each frequency. Blue arrows indicate Eu spin flip processes, red arrows indicate Y spin flips, and purple arrows indicate Y flip-flop processes. Inset: plotted frequencies without color scale.}
\label{fig:dataEu}
\end{figure*}

A 1000~ppm $\mathrm{Eu:YSO}$ crystal is cooled to 4~K in a closed-cycle cryostat. Population manipulations are performed to prepare a 3~MHz wide atomic frequency comb on the $\ket{\pm5/2}_g\leftrightarrow\ket{\pm1/2}_e$ transition with a preparation beam (gray Prep beam in Fig.~\ref{fig:Euexp_setup}), allowing to reach a double-pass optical depth of 6. Then the spin-wave storage of classical pulses of light is performed in the ensemble, by sending an optical input pulse on the $\ket{\pm5/2}_g\leftrightarrow\ket{\pm1/2}_e$ transition (orange In beam in Fig.~\ref{fig:Euexp_setup}), followed by a 15~$\mu$s-long optical control pulse on the $\ket{\pm3/2}_g\leftrightarrow\ket{\pm1/2}_e$ transition (blue Ctrl beam in Fig.~\ref{fig:Euexp_setup}). Note that all beams propagate along a direction close to the $b$ axis (offset angles serve for spatial filtering) and are all polarized along $D_1$. Spin rephasing is then performed by applying an RF XY-4 sequence on the 46.2~MHz $\ket{\pm5/2}_g\leftrightarrow\ket{\pm3/2}_g$ transition. The RF field is applied thanks to a coil wrapped around the crystal. Details about pulse shape, preparation procedure and pulse sequences can be found in \cite{ortu_storage_2022} and its Sup. Mat., for which the same experimental setup and sequence was used.
An additional static magnetic field of varying amplitude is applied along $D_{1}$. It is important to mention that in the experimental sequence, the RF pulses compensate for the effect of the inhomogeneous spin decoherence, such that the plotted decay curves are limited by single-ion decoherence, which can be compared to the model discussed in this paper.\\
The decays of the spin-wave storage of classical pulses of light are presented with orange dots in Fig.~\ref{fig:Eu_echoes} for bias magnetic field amplitudes of 1, 10, 40 and 60~$\mu$T. These curves display a clear dependency on the applied external field, as well as modulation patterns.\\
To simulate this behavior, we use the previously introduced model and sequence, with an hypothesis on the initial coherence creation pulse: in the experimental case, the initial coherence is indeed created through optical excitation followed by an optical transfer pulse, each of the two pulses having very different parameters. This is in contrast with our simulations, where all calculations and simulations are done on the spin transitions. Therefore, we chose to mimic this by using a 200~$\mu$s FWHM adiabatic pulse on the spin transition, chirped by $50~$kHz. Such a choice essentially conditions the depth of the modulations at the Zeeman frequency but not the overall shape of the coherence decay~\cite{PhysRevA.82.042309,PhysRevA.69.022321}. The simulated curves are shown in blue, by considering the first five Y neighbors. Even if the simulated sequence is much simpler than the actual full spin-wave protocol, we can observe two important correspondences.\\
The first agreement is the global decay time, which is well reproduced: an increase in the amplitude of the magnetic field leads to a longer coherence time, from less than a ms at low field to 5~ms at 60~$\mu$T. Here we remind that no empirical decay is introduced manually in the shape of the coherence decay~(\ref{analyt_dyn}) (no $e^{-(t/\tau)^x}$ term introduced): it is solely due to path interference in the $\hat{H}_0$ manifold driven by the $\hat{H}_{\rm int}$ Hamiltonian. Therefore, our simulations always present long-duration coherence revivals, that are pushed further and further by considering more and more bath ions.

The second observation is the faster modulation present in the dynamics, at a frequency of the Eu Zeeman splitting ($\approx 14~$kHz/mT), that we also reproduce numerically. The contrast of this modulation also diminishes when increasing the magnetic field amplitude. This is mainly due to the decrease of the probability of having a central spin flip during the coherence creation with an increasing field, as we showed in a previous publication~\cite{PhysRevA.103.022618} (both Zeeman splittings of the spin transitions are actually involved in the evolution). Moreover, the experimental decay pattern strongly depends on the rephasing pulse position, that we do not take into account due to the simplified sequence considered here~\cite{PhysRevA.103.022618}. It can also be observed that the mismatch between the curves is higher at low magnetic fields, which we attribute to a difficulty to precisely estimate experimentally the applied magnetic field for such low values.
For quick comparison with the first-order CCE approximation, we also show the curves obtained with the first 30 Y neighbors with black dashed lines in Fig.~\ref{fig:Eu_echoes}. The simulated CCE curves clearly display longer decays, as they do not grasp the main decoherence mechanisms for this RE, as we will see below.\par

It should also be pointed that in the experimental echo decays, one revival stands precisely at the yttrium Zeeman frequency (3.6~ms at $40~\mu$T and $6~$ms at $60~\mu$T), which is not reproduced by our simulations. This discrepancy leads us to investigate more in details the different frequencies involved in the dynamics, as well as their relative weights. Figure~\ref{fig:2Y_Eu_freq_weights} shows the frequencies $\Delta\lambda_{l,l'}$ and their respective frequency weights $g_{l,l'}$ of Eq.~(\ref{analyt_dyn}) (color scale), as a function of the magnetic field amplitude between 0 and 60~$\mu$T. In the inset, all the dynamical frequencies $\Delta\lambda_{l,l'}$ are plotted without the color scale. The frequency weights show the different processes at play depending on the magnetic field amplitude. Up to $10~\mu$T, all Eu spin flip (blue arrows), Y spin flip (red arrows), and Y flip-flop processes (purple arrows) are non-negligible. Beyond $10~\mu$T and up to $40~\mu$T, as the frequency gap between the Eu Zeeman level increases, the Eu-Y dipole-dipole interactions are not sufficient to induce Eu spin-flips anymore. Above $40~\mu$T, Y spin-flips are forbidden for the same reason, and only Y flip-flops contribute. Among these mechanisms, only the Y spin flips are taken into account in the first-order CCE, explaining the previously mentioned mismatch.\par

Above $40~\mu$T, predominant dynamical frequencies stay mostly constant with the field, leading to a constant characteristic time in the echo decays (see Fig.~\ref{fig:Eu_echoes}). It should also be pointed that Fig.~\ref{fig:2Y_Eu_freq_weights} is plotted with two Y ions, whereas the echo decays of Fig.~\ref{fig:Eu_echoes} are simulated with five Y ions. Therefore the actual amount of dynamical frequencies giving rise to the decay is higher than the one shown here. However, this graph indicates which process is relevant in each field regime. To further strengthen this interpretation, we propose in App.~\ref{app:pop_field} to show population dynamics that reveal the aforementioned spin flips.\\
The last necessary validation concerns the pertinence of the approach consisting in simulating a simple FID process with respect to a full echo sequence. As mentioned in Sec.~\ref{subsec_approx}, the $\hat{A}$ term in the dipole-dipole coupling creates a level-dependent energy shift: if the system is in an initial mixed state, it will dephase quickly as it is the case for an inhomogeneously broadened medium. This occurs even if we consider a single ion interacting with the bath. Therefore, we should ensure that the limiting decay time (essentially independent of the magnetic field) is long enough as compared to the simulated one. To this end, we plot in Fig.~\ref{fig:Eu_echoes} the decays associated with Loschmidt echoes \cite{Rossini08}, in which static decoherence is not compensated by a rephasing pulse as in either Hahn echo or spin-wave AFC sequences. The curves are represented in solid black. They reveal that the simulated coherence decay begins to be similar to Loschmidt echos above 40~$\mu$T, showing that simulation at higher fields would require to introduce rephasing pulses that also counteract single-ion inhomogeneous dephasing. We confirm the validity of our approach at fields up to 60~$\mu$T in App.~\ref{app:Hahn} by comparing a Hahn echo sequence with our simulated sequence, and witnessing a good correspondence.
\begin{figure*}
\subfloat[\label{fig:Y_echoes}]{%
  \includegraphics[width=1.2\columnwidth]{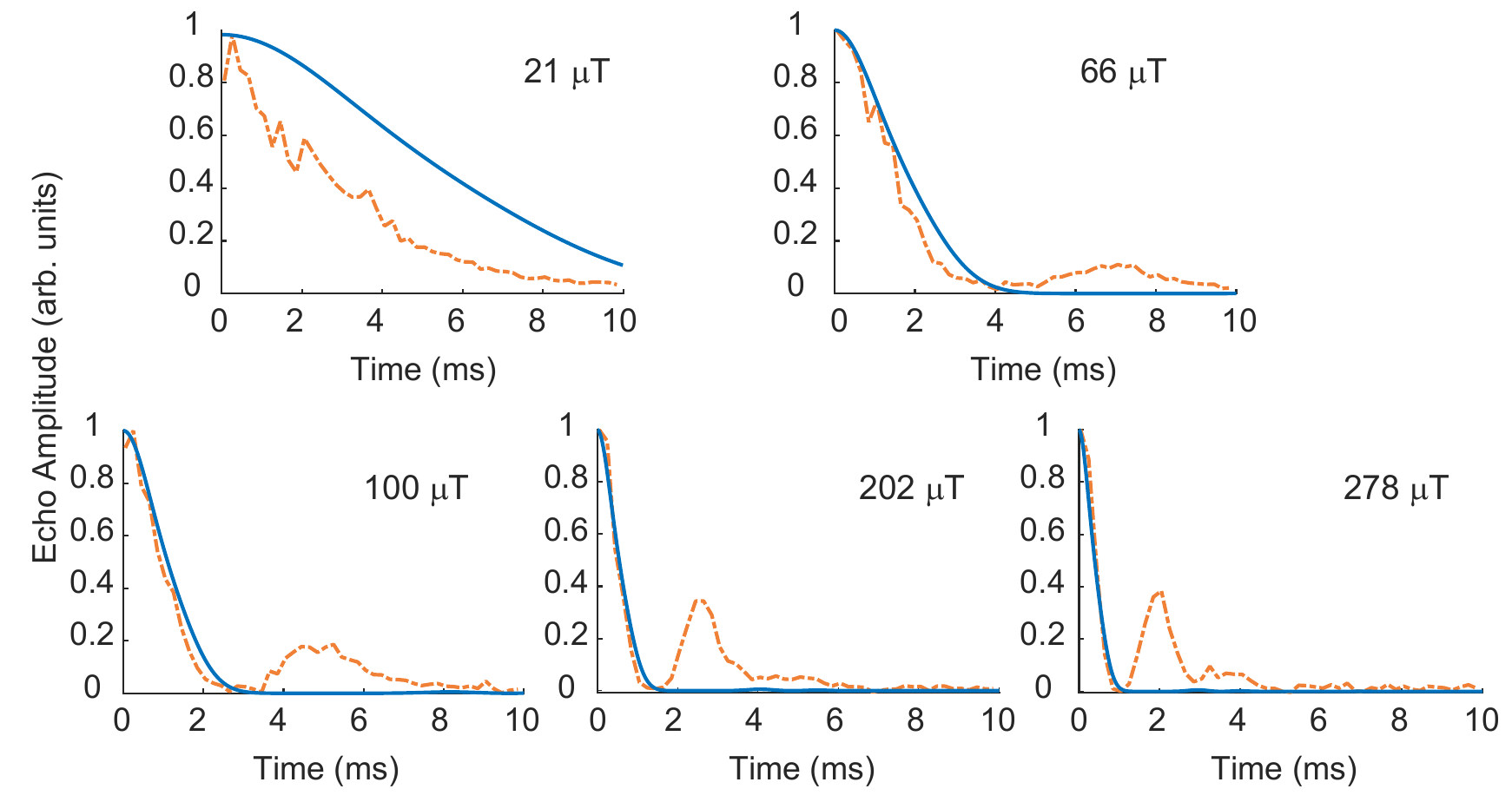}%
}\hfill
\subfloat[\label{fig:2Y_Yb_freq_weights}]{%
  \includegraphics[width=0.8\columnwidth]{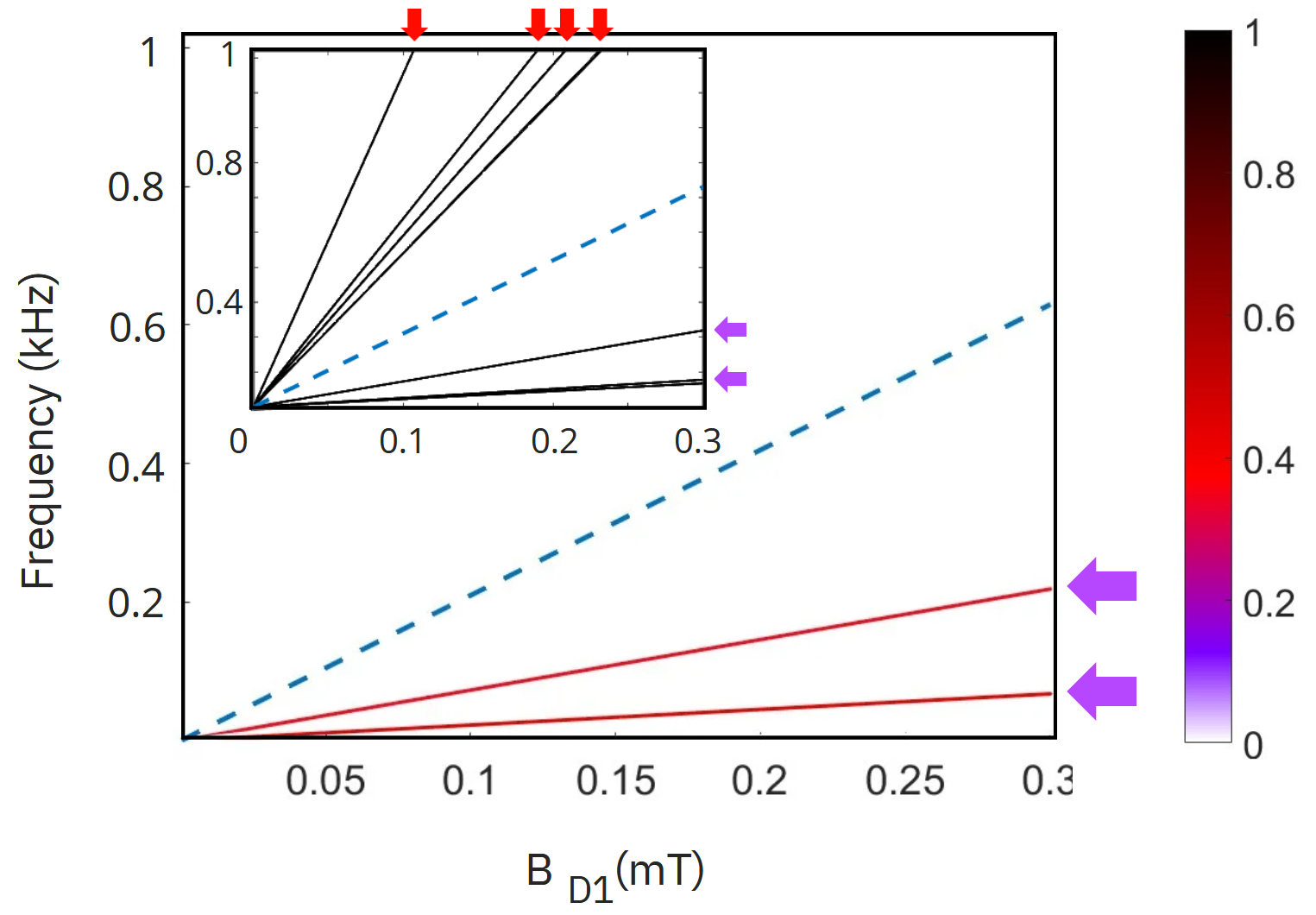}%
}\hfill
\caption{a) Yb echo decays at various magnetic field amplitudes. Simulations for one Yb ion and 7 Y ions (in blue) and experimental data (in orange) \cite{Nicolas2022}.  b) Dynamical frequencies involved in the echo dynamics as a function of the magnetic field amplitude. The color scale represents the $g$ weight (see Eq. (\ref{analyt_dyn})) associated with each frequency (normalized). Inset: frequencies involved in the echo dynamics, without color scaling.  The red arrows point out components associated with Y spin flips processes, and purple arrows points components associated with Y flip-flops processes. In both figures, the blue dashed line represents the $2.1~$kHz/mT Y Zeeman frequency found in the experimental echo decay revivals of Fig.~\ref{fig:Y_echoes}.}
\label{fig:gatherYb}
\end{figure*}
\subsection{Ytterbium}\label{Yb_sec}

The second RE ion investigated is ytterbium ($\mathrm{Yb:YSO}$). Recent studies have revealed long coherence times in particular magnetic configurations \cite{ortu_simultaneous_2018,Nicolas2022}, and we propose here to confront our simulations with previously published experimental echo decays. Experimental data points and details about the used sequence can all be found in~\cite{Nicolas2022}. The experimental measurement of the coherence is performed using a Hahn echo sequence and detected through coherent RHS techniques. Both the experiments and our simulations are performed on the $2.497$~GHz transition as stated before, for the optical site II (see Fig.~\ref{fig:data1}).\\
In our first simulations, we aim at replicating the experimental echo decays recorded at varying magnetic field amplitude along the $D_1$ direction shown in Fig.~\ref{fig:Y_echoes} in orange.
We perform our simulations with seven Y ions and the decays are represented in blue in the figure. For magnetic fields above $\sim 60~\mu $T, we obtain a good correspondence in terms of initial decay time between the curves. For magnetic field amplitudes below $60~\mu $T, the simulation predicts a longer decay time than experimentally measured. 
This difference can be explained by the difficulty to precisely estimate the magnetic field experimentally (in the same way as for europium in the previous section), and by the fact that in this regime the superhyperfine interaction might not be the limiting process anymore.
In addition to the overall good match in the decay characteristic time, we also notice that the initial horizontal slope is well reproduced by our simulations. This is explained by the fact that in our approach, the decay is composed of multiple sinusoidal contributions, each with a zero slope at zero delay. This results in Gaussian-like decays, that are in good agreement with experimental data, as illustrated in Fig.~\ref{fig:Y_echoes}.\par
\begin{figure}[h]
  \includegraphics[width=0.8\linewidth]{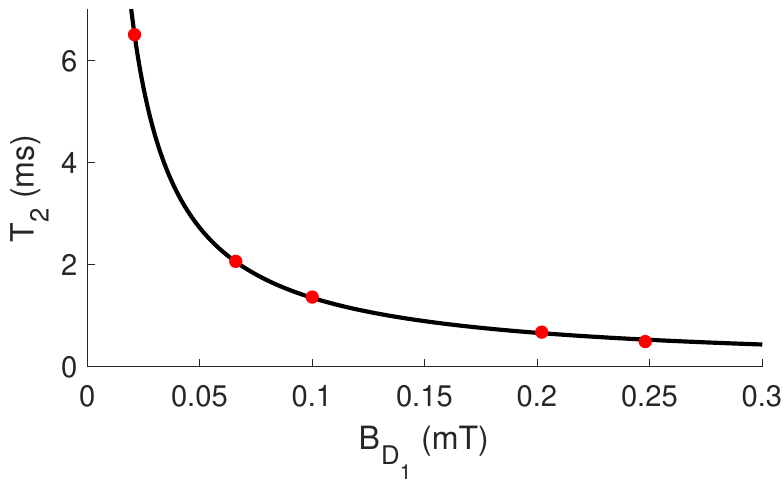}
  \caption{Dependence of the simulated decay time in Yb:YSO as a function of the magnetic field amplitude. $T_2$ red dots values fitted from the simulated coherence decays of Fig.~\ref{fig:Y_echoes}. Black line is a fit of red dots, with a law $T_2=A_1/B_{D_1}$ and $A_1=1.36\cdot10^{-7}$~s.T. }
  \label{fig:1sBdep}
\end{figure}

\begin{figure*}
\subfloat[\label{subfig:71}]{%
  \includegraphics[width=0.62\columnwidth]{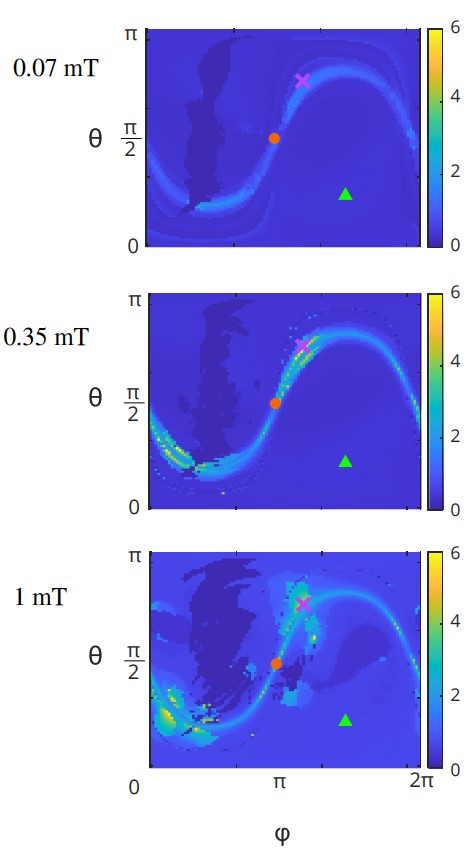}%
}\hfill
\subfloat[ \label{subfig:72}]{%
  \includegraphics[width=0.511\columnwidth]{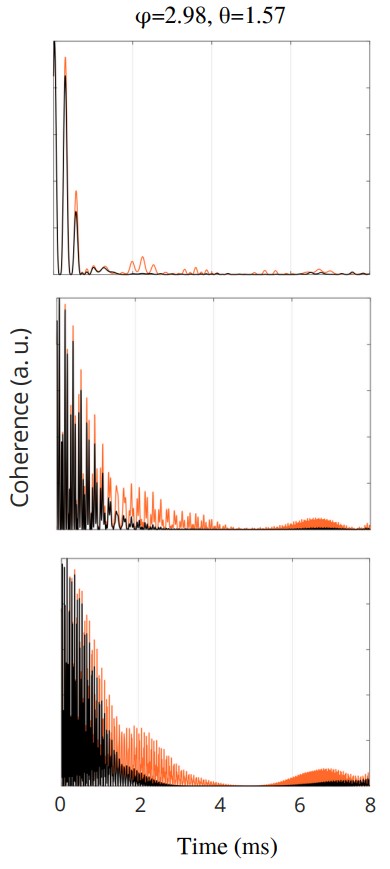}%
}\hfill
\subfloat[ \label{subfig:73}]{%
  \includegraphics[width=0.469\columnwidth]{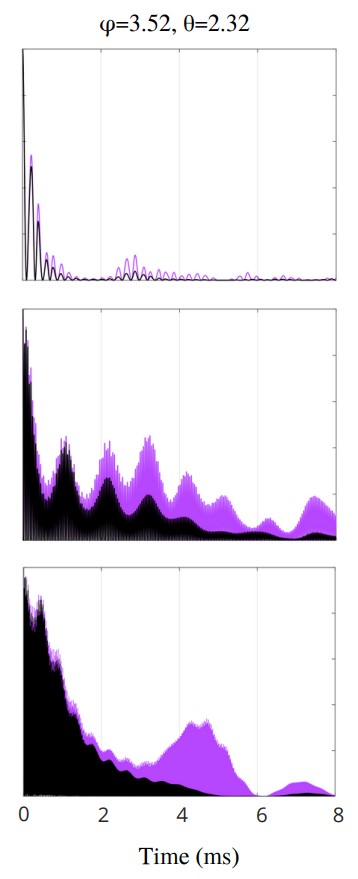}%
}\hfill
\subfloat[ \label{subfig:74}]{%
  \includegraphics[width=0.473\columnwidth]{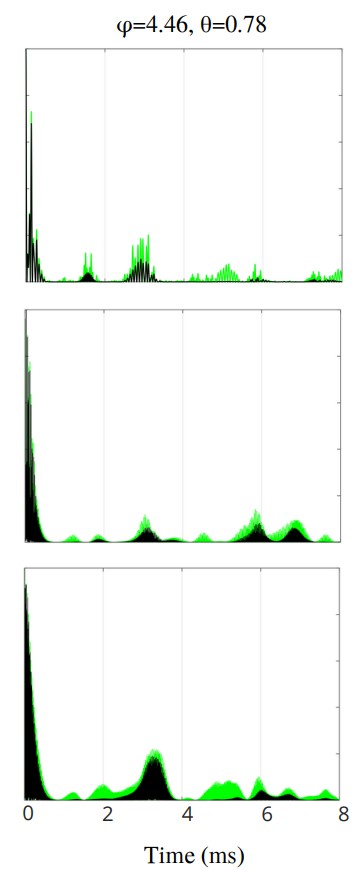}%
}\hfill
\caption{ a) Simulated coherence time (color map, in ms) as a function of magnetic field angles $\varphi$ and $\theta$ (see Eq.~(\ref{B})) for one Pr ion and three Y ions and for magnetic field amplitudes of $B_{\rm DC}=$~0.07, 0.35 and 1~mT. Simulations are performed for one magnetic subsite. Echo decays associated with colored symbols in Fig. (a) are represented on: b) for the dot (orange), c) for the cross (purple), and d) for the triangle (green). For comparison, all echo decay simulations were also performed with five Y ions (black curves).}
\label{fig:data7}
\end{figure*}
The second point highlighted by these curves is the absence of revivals in the simulated decays as compared to experimental ones, in which a clear revival close to the Y Zeeman splitting uncoupled frequency ($2.1~$kHz/mT) is visible. Two possible explanations are proposed. On one hand, the mismatch could come from the initial state, that may not be in reality a perfectly balanced mixed state of the levels inside the Y manifold (see the $\alpha_k(0)\alpha_k'(0)$ term in Eq.~(\ref{analyt_dyn})). The consequence for this is that different frequency components involved in the dynamics can interfere in a different way given their different weights.
On the other hand, we do not simulate an optical pulse to transfer the coherence from one ground hyperfine Yb level to an excited hyperfine level. Instead we assume that we have a perfect population transfer from the $\ket{2}_{g}$ state to the $\ket{2}_{e}$ state, similarly as for europium (see Sec.~\ref{Eu_sec}). This idealized population transfer in our simulations may also affect the dynamics. \par

Then, similarly as for Eu, we associate the frequencies observed in the dynamics with spin processes. To this extent, we plot in Fig.~\ref{fig:2Y_Yb_freq_weights} the evolution of the dynamical frequencies $\Delta\lambda_{l,l'}$ as a function of the magnetic field amplitude for one Yb ion and two Y ions, still for the same field orientation (aligned along $D_1$). Their respective weights $g$  are normalized and represented with a color scale. The Y Zeeman frequency $2.1~$kHz/mT is plotted as a blue dashed line, and it clearly appears that such a frequency is actually not present in the spin dynamics, corroborating the observations made with the simulated decays in Fig.~\ref{fig:Y_echoes}. Frequencies with the strongest weights stand at lower frequencies, at $0.67~$kHz/mT and $0.3~$kHz/mT. To associate each curve with a physical process, we plot in the inset of Fig.~\ref{fig:2Y_Yb_freq_weights} all dynamical frequencies (unweighted, as in Fig.~\ref{fig:2Y_Eu_freq_weights}). In this inset, three groups of curves can be identified. The highest energy branches correspond to double Y spin flips (2 branches that are indistinguishable on the figure), then a group of 8 branches (3 visible on the figure) correspond to single Y spin flips. Eventually, at lower energies, 4 branches (3 visible on the figure) are associated with Y flip-flops. Notice here that these branches are actually field-dependent due to the linear dependence of the magnetic dipole of Yb with the field, on the contrary to the previous section with Eu. This feature is addressed in detail in the Supp. Mat. of~\cite{Nicolas2022}. In our simulated decays, we see that the only components with non-negligible weights are the Y flip-flop components, driven at a frequency essentially governed by the Yb-Y dipole-dipole coupling (one order of magnitude larger than simple Y-Y dipole-dipole interaction).\\
Another important difference of this system with Eu is the constant weights of the involved dynamical frequencies, because of the linearly increasing magnetic dipole of Yb with the field. This implies that augmenting the field amplitude increases the dynamical frequencies at stake while maintaining their weight constant. The consequence for this is that the frequency map remains identical whatever the field up to a global scaling factor. In other words, changing the field by a factor $K$ amounts to changing time by a factor $1/K$. To quantify this effect more explicitly, we extract from our four simulated coherence decay curves (Fig.~\ref{fig:Y_echoes}) a coherence time $T_2$ by fitting them with a stretched exponential $\exp[-(t/T_2)^\beta]$. The resulting $T_2$ are plotted in Fig.~\ref{fig:1sBdep} with red dots and are well fitted with a $1/B$ scaling law, as illustrated by the black solid line.\par

\subsection{Praseodymium}\label{Pr_sec}

\begin{figure*}
\subfloat[\label{fig:Pr_Y_Ze}]{%
  \includegraphics[width=0.58\columnwidth]{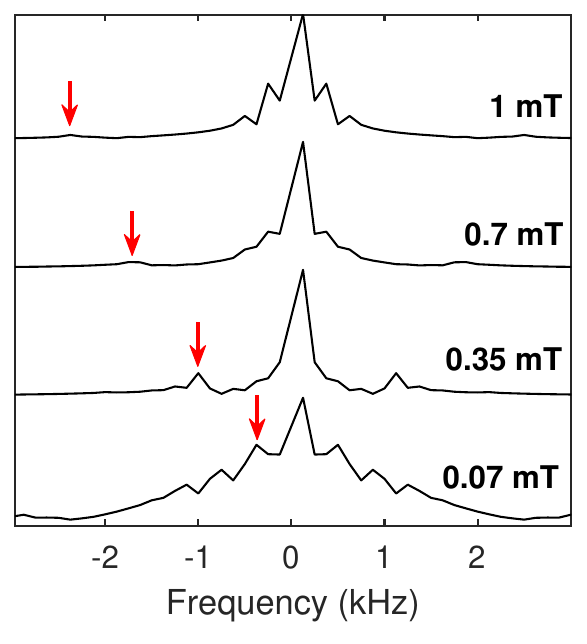}%
}\hfill
\subfloat[ \label{fig:S1_pr}]{%
  \includegraphics[width=0.7\columnwidth]{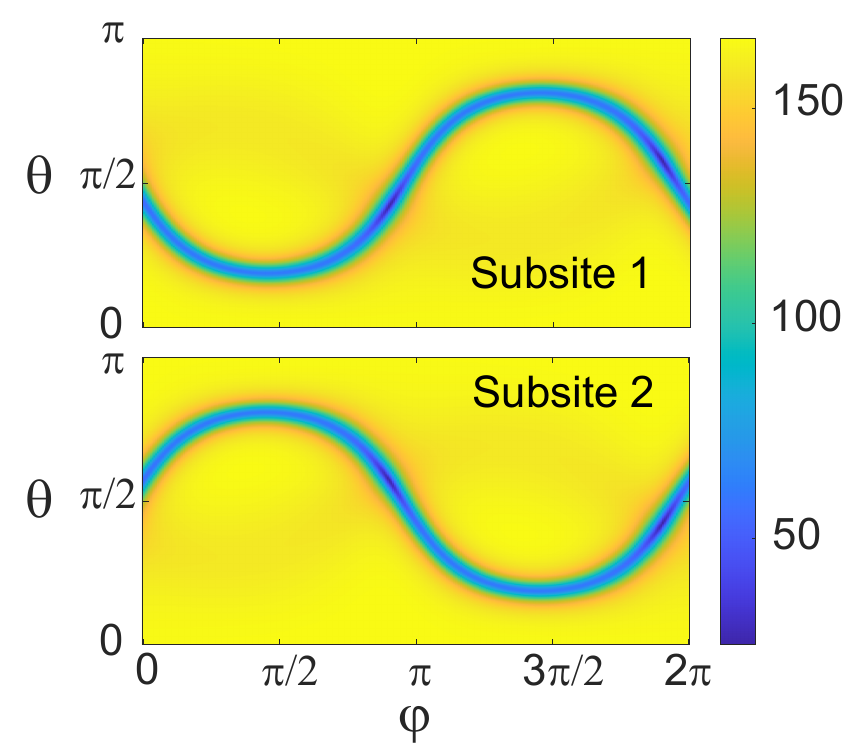}%
}\hfill
\subfloat[ \label{fig:3Y_Pr_D1D2_cohmap_zoom}]{%
  \includegraphics[width=0.75\columnwidth]{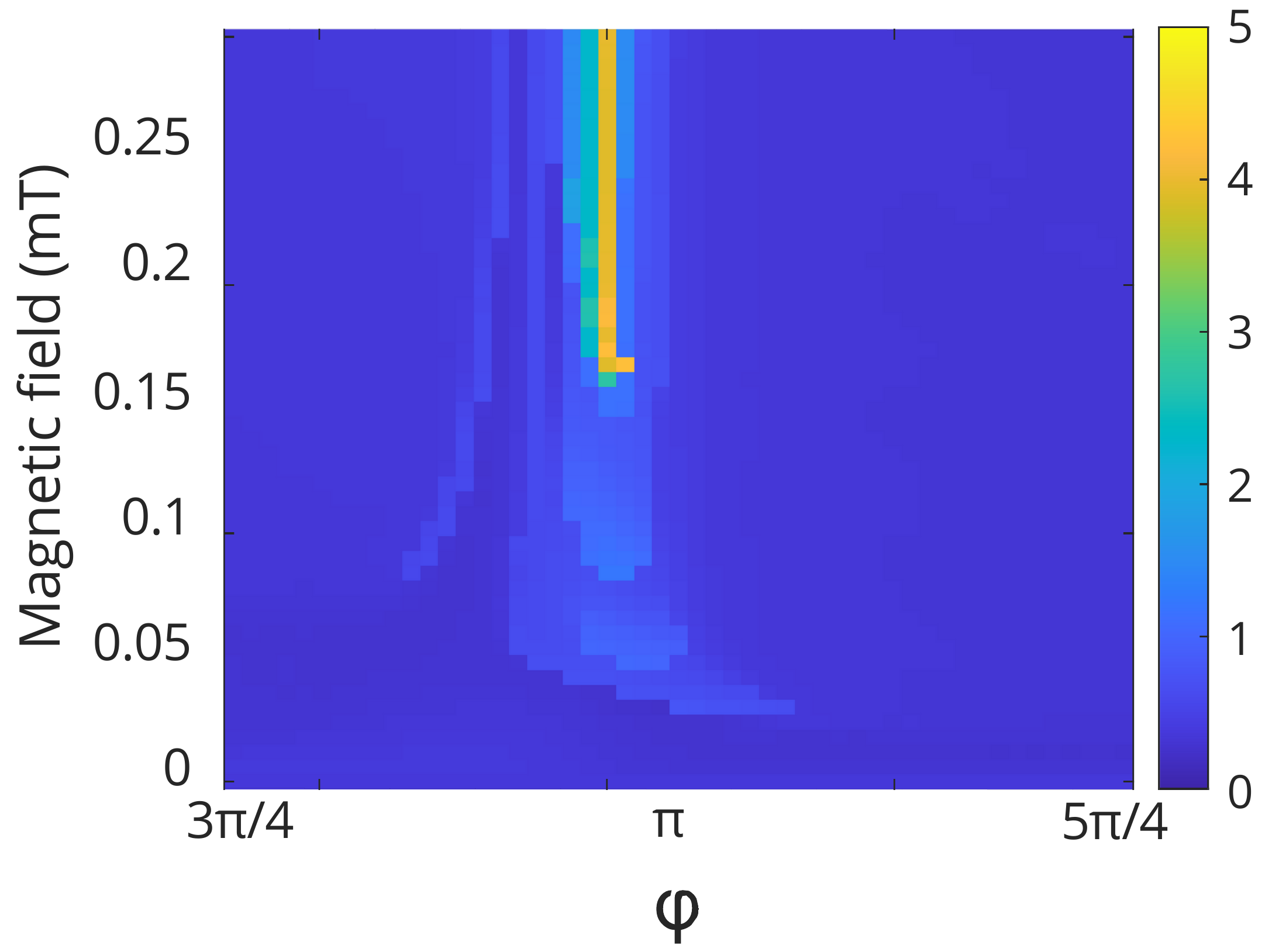}%
}\hfill
\caption{ a) Fourier transform of the echo decays presented in Fig.~\ref{subfig:73}. The red arrows indicate the frequencies corresponding to the $2.1~$kHz/mT yttrium nuclear magnetic moment. b) Mean $||\mathbf{S}_{1}||$ (MHz/T) of the four transitions $\ket{\pm1/2}_g\leftrightarrow\ket{\pm3/2}_g$ for each magnetic subsite as a function of the magnetic field direction, for an amplitude of 1~mT. c) Coherence time (color map, in ms) as a function of magnetic field angle $\varphi$ (in rad) and amplitude for one Pr ion and three Y ions in the ($D_1$,$D_2$) plane.}
\label{fig:pr}
\end{figure*}

In this section, we will show numerical predictions of spin coherence times for Pr:YSO (see Fig.~\ref{fig:data1} for the energetic structure). As stated in Sec.~\ref{ex_model}, we will focus on the spin transition $\ket{\pm 1/2}_g\leftrightarrow\ket{\pm 3/2}_g$ at 10.2 MHz. \par
We plot in Fig.~\ref{subfig:71} the coherence time as a function of the magnetic field angles $\varphi\in[0,2\pi]$ and $\theta\in[0,\pi]$ (see Eq.~(\ref{B}) for angles definition), for magnetic field amplitudes of $B_{\rm DC}=$~0.07, 0.35 and 1~mT. In this graph, due to simulation duration constraints, we limit the number of Y ions in the system to 3, and the simulation time range to 6.5~ms. We recall that the coherence time is inferred with the method presented in App.~\ref{App_A}. The first information revealed by these plots is the high anisotropy of the coherence time with the field orientation: at magnetic fields above 0.35~mT, there is a factor of ten between low and high coherence time regions ($\approx$~0.5 to 5~ms). This feature is to be linked with the high anisotropy of the magnetic dipole of Pr:YSO, with a factor of more than 4 difference along the three directions of space~\cite{PhysRevLett.92.077601}. The second information is that the coherence time increases when the magnetic field amplitude (up to 0.35~mT here), as witnessed experimentally~\cite{Fraval2004b,Fraval2004a}. However, the plots show that above this value, increasing the magnetic field does not lead to a coherence time extension anymore. This observation is in agreement with simulations performed with Eu:YSO in Sec.~\ref{Eu_sec}, in the regime where Y spin flip-flops are the only remaining decoherence mechanism: both REs have a field amplitude independent magnetic dipole (as long as the quadratic Zeeman contribution remains negligible).\\
In order to further investigate the improvement in the coherence time, we plot the decays in three configurations indicated by colored symbols in Fig.~\ref{subfig:71}. These are represented in Fig.~\ref{subfig:72} for the orange dot, Fig.~\ref{subfig:73} for the purple cross, and Fig.~\ref{subfig:74} for the green triangle. Notice that for these particular curves, we push the simulation duration up to 8~ms to visualize the time dependence in more details. The curves reveal the non-monotonous nature of the decay and display, similarly as for Eu:YSO (see Sec.~\ref{Eu_sec}), a strong component at the Pr Zeeman frequencies. In the same manner, the depth of the modulation is directly conditioned by the initial state, that we however don't aim at adjusting here.\\
Then, in order to assess if limiting the simulation to the first three neighboring Y ions is sufficient, we plot the echoes obtained with five ions in black on top of the three Y ions curves. Adding more Y ions does not change the hierarchy of decay times (highest coherence regions identified in Fig.~\ref{subfig:71} still lead to the longest coherence times with 5 Y ions) but modifies the decay shape. In particular, strong revivals appearing after a few milliseconds are blurred by the introduction of additional accessible energies in the system. Notice that this is also the case for Yb:YSO, in which strong revival can appear in decay simulations if the number of surrounding Y taken into account is insufficient.\par
In Fig.~\ref{subfig:73}, we also clearly identify beatings at the yttrium Zeeman frequency (particularly visible at 0.35~mT). The frequency components of the corresponding decays are shown in Fig~\ref{fig:Pr_Y_Ze}, and correspond to spectral structures that have also been measured experimentally in Pr:YSO~\cite{Fraval2004b}, and associated with single or multiple Y spin flips by the authors. We confirm here this interpretation numerically.\par

Finally, in order to gain insight into the shape of the region of long coherence times (sinusoidal like shape in Fig.~\ref{subfig:71}), we plot the average of the norm of the four gradients $||\mathbf{S}_{1}||$ of the transitions involved in the four distinct spin transitions 
$\ket{\pm1/2}_g\leftrightarrow\ket{\pm3/2}_g$ in Fig.~\ref{fig:S1_pr}. In this plot, we consider both magnetic subsites when calculating the average $||\mathbf{S}_{1}||$ because when $\theta\notin\{0,\pi/2\}$, both subsites are nonequivalent. The concordance of the shapes for subsite 1 shows that the sinusoidal pattern indeed mostly originates from a weaker sensitivity of the transitions to external fluctuations. However, the longest coherence times simulated in Fig.~\ref{subfig:71} are not always superimposed with the minimum of $||\mathbf{S}_{1}||$, indicating that a better optimization of the coherence time could be achieved with our model. Experimentally, we expect the decays to consist of contributions of both subsites. Therefore, a simple working point to be identified in these plots is at the crossing of the two low $||\mathbf{S}_{1}||$ zones in Fig.~\ref{fig:S1_pr}. This point corresponds to $\theta=\pi/2$ (field in the ($D_1$,$D_2$) plane), where the two magnetic subsites are equivalent by symmetry, with a corresponding decay plotted in Fig.~\ref{subfig:72}. We focus on this region by also plotting the coherence time as a function of $\varphi$ and the field amplitude in Fig.~\ref{fig:3Y_Pr_D1D2_cohmap_zoom}, showing that it is for $\varphi$ close to $\pi$ (field along $D_1$) that the coherence time is the highest. It also gives the value of the field ($\sim200~\mu$T) for which Y spin flip-flops remain the only decoherence mechanism. This is five times higher than for Eu, in agreement with the strength of the RE-Y dipole-dipole coupling as we will see in the next section.

\section{Discussion}
\subsection{Impact of the Y ions on the coherence times}\label{Y_sec}

Both the simulations and the experimental data presented previously showed that the coherence times in RE-Y systems are highly dependent on the magnetic field amplitude and orientation, as witnessed for Pr in Fig.~\ref{subfig:71} and for Yb in App.~\ref{app:Ybcoh}, Fig.~\ref{fig:3Y_Yb_2D_map}. Due to the anisotropy of the magnetic moments of the RE ions, as the magnetic field orientation changes, so does the strength of the RE ion effective magnetic dipole and consequently, the dipole-dipole RE-Y interaction strength.\par

In the case of Yb, as the magnetic dipole linearly increases with the applied magnetic field, the coupling with the surrounding Y ions increases accordingly. This leads to the contribution of dynamical frequencies of increasing frequencies with the field, with weights that do not vanish when increasing the amplitude of the field~\cite{Nicolas2022}.
On the contrary, for a given magnetic field orientation, the magnetic dipoles of the Eu and Pr ions are constant regardless of the magnetic field amplitude. The constant strength of the Eu-Y and Pr-Y dipole-dipole interactions result in a different impact on the whole RE-Y dynamics, depending on the field amplitude. This leads us to identify three magnetic field regimes for the Eu and Pr ions.\par
\paragraph{Very low field regime: }
When both the RE Zeeman splitting $\delta_{RE}$ and the Y Zeeman splitting $\delta_{Y}$ are smaller than the RE-Y magnetic dipole-dipole coupling, we are in the very low field regime. In this regime, RE-Y interactions can cause RE spin flips (blue arrows in Fig.~\ref{fig:data1}), Y spin flips (red arrows in Fig.~\ref{fig:data1}) and Y flip-flops (purple arrows in Fig.~\ref{fig:data1}). This causes the dynamics of the system to present many frequencies with non-negligible weights $g$ (see Eq.(\ref{analyt_dyn})), which may also span a larger frequency range than for higher fields. Both effects lead to fast decays, and are observed in Fig.~\ref{fig:dataEu} for Eu, for magnetic field amplitudes below $10~\mu$T.\par
\paragraph{Low field regime: } When the Y Zeeman splittings $\delta_{Y}$ are of the same order of magnitude than the dipole-dipole RE-Y coupling, we enter the low magnetic field regime where the RE-Y couplings are not able to induce RE spin flips anymore. We saw that this regime is situated between $10~\mu$T and $40~\mu$T for Eu (field along $D_{1}$).
\paragraph{Intermediate field regime: }Finally, when both $\delta_{RE}$ and $\delta_{Y}$ are larger than the RE-Y coupling, we enter the intermediate magnetic field regime. Dipole-dipole couplings are not able to cause any spin flips but can only drive Y flip-flops. If the magnetic dipole of the RE ion does not scale with the magnetic field amplitude, such as for Eu and Pr, the coherence time will become mostly independent on the magnetic field amplitude. Indeed, the frequency splittings created by the dipole-dipole interactions are independent of the field amplitude, and consequently the spectral range explored in the system dynamics will remain unchanged. This spectral range being narrower than the spectral range explored in low and very low magnetic field regimes, the intermediate magnetic field regime is the one that provides the longest coherence times among the three regimes investigated here.\\

\subsection{Perspectives}\label{persp_sec}

\begin{figure}
  \includegraphics[width=0.9\columnwidth]{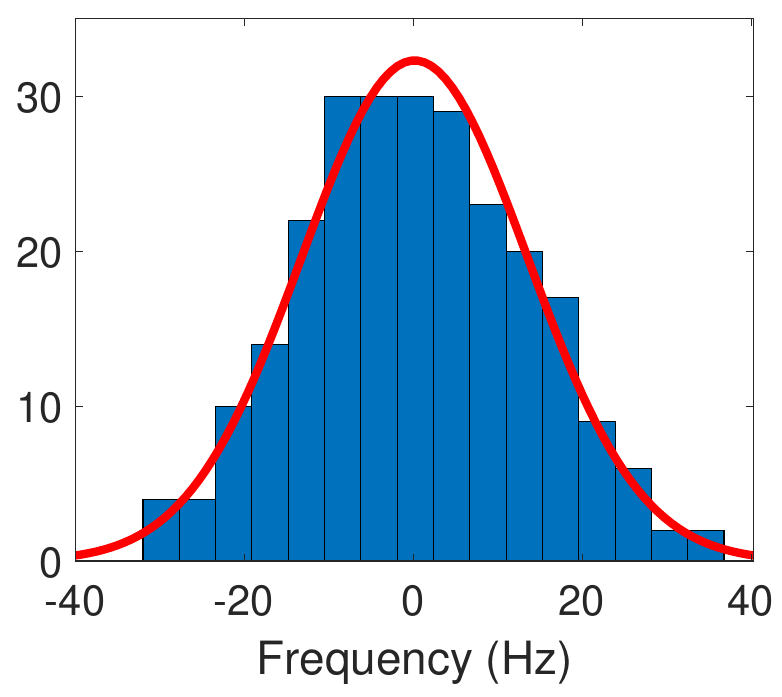}%
\caption{Histogram of the energy distribution of the $Iz=0$ manifold of the first ten Y ions (blue histogram, occurrence in the vertical axis). The energy spread is due to Y-Y magnetic dipole coupling (see second term in Eq.~(\ref{bath_Hamilton})), and fits with a Gaussian of FWHM 30~Hz (solid red).}
\label{spread}
\end{figure}

Our work has allowed us to identify decoherence mechanisms in different magnetic field regimes in RE-doped YSO. Simulations have helped to identify favorable regimes, by mainly focusing on fields applied in the ($D_1$,$D_2$) plane for symmetry reasons. A deeper investigation of the the interplay between the two magnetic subsites could potentially help pushing the coherence times even further (see for instance high coherence time zones of subsite 1 in Fig.~\ref{fig:data7}).\\
The sequence that we have used to simulate the coherence times shall also be discussed: in the very low field regime, only a few hypotheses can be done on the total Hamiltonian, therefore full dynamics simulation must be done for estimating the coherence time. Limited computation power ultimately limits the complexity of the sequences that are simulated, as well as the size of the Hilbert space (and therefore the number of Y ions). In the very low field regime, we have seen that the coherence decays could reproduce well the experimental echo decays, indicating that taking into account only a few Y ions is sufficient. However, we predict that if one pushes the field further, the simplistic sequence used here will not reflect anymore actual coherence times. This is mainly due to the single-ion z-coupling contribution discussed at the end of Sec.~\ref{Eu_sec}. However, entering the intermediate field regimes also means that spin flips are forbidden, and therefore that frozen central spin approximation can be made. Given that first-order CCE does not allow to simulate ion bath flip-flop, we believe that higher-order CCE would be a viable and efficient way of simulating the system in such regime. It would also allow to focus more attention on the spin manipulation during spin rephasing sequences: imperfections in spin inversions can cause the rephasing sequence to be less efficient than expected~\cite{PhysRevA.103.022618}, therefore a precise determination of the magnetic field regime and inversion pulse parameter is still to be precisely investigated.\\
Such investigation would also allow us to simulate more complex sequences such as dynamical decoupling sequences, routinely applied to counteract spectral diffusion, that is usually attributed to Y flip-flops in the intermediate field regime~\cite{ortu_storage_2022, Holzäpfel_2020, PhysRevB.86.184301}. In this regime, it is therefore instructive to illustrate the range of frequencies involved in such flip-flop dynamics. In our picture, we recall that the Y flip-flop mechanism gives rise to a lift of degeneracy of equal total Y spin projection states (see Fig.~\ref{fig:data1}). If we plot the spread of frequencies spanned by the distribution of the first ten Y nearest neighbors, we obtain the histogram of Fig.~\ref{spread} for the configuration 5 spins "up" and 5 spin "down", leading to a total of C$^5_{10}=252$ configurations. This distribution fits well with a Gaussian distribution with FWHM of 30 Hz. Links between this energy spread and spectral diffusion witnessed in non-Kramers ions deserve further investigation, but such a value is already in good agreement with typical linewidths considered in spectral diffusion models (35~Hz found for instance in~\cite{Holzäpfel_2020}).

By pushing the field amplitude further, the quadratic Zeeman contribution (third term in Eq.~(\ref{eq:nonKramersH0})) becomes non negligible anymore, and we believe that our approach can provide new paths for the comprehension of mechanisms such as the `frozen core' effect~\cite{PhysRevLett.92.077601,Zhong2015}.

\section{Conclusion}
By performing a small-scale exact treatment of the spin interactions in RE ion-doped crystals in the $\mu$T to mT magnetic field amplitude range, we have shown that we could understand and simulate the processes that induce decoherence and experimentally limit the coherence times. We have validated our approach with a side-to-side comparison between simulations and experimental results for both Eu and Yb, and although slight differences in the dynamics have been witnessed, similar coherence decay times have been found in the magnetic field amplitude range of 0-100~$\mu$T, with no adjustable parameter. Remarkably, in Yb:YSO the $1/B$ dependency of the experimental echo decay time could be well reproduced numerically. Following this, we have used our simulator to make coherence time predictions as a function of magnetic field amplitude and orientation in a Pr:YSO system, and have identified a favorable set of magnetic field parameters that minimize the decoherence induced by the surrounding Y ions.\par

Our results have led to the identification of different magnetic field regimes, in which different spin flip processes induced by the RE-Y dipole-dipole interactions are predominant.

\acknowledgments{The authors thank A. Holz\"apfel, H. de Riedmatten, F. Appas, P. Goldner and A. Tiranov for fruitful discussions and important feedbacks on the simulations. JE acknowledges the National Research Agency (ANR) through the project WAQUAM (ANR-21-CE47-0001-01), France 2030 National Research Program on Quantum Technologies project Qmemo (ANR-22-PETQ-0010), Univ. Côte d'Azur CSI through the project FITRALP, Doeblin Federation through the project STALAR and DEP from Académie 2 Univ. Côte d'Azur. VDA acknowledges the Institut Universitaire de France. MA acknowledges funding from the Swiss FNS NCCR program Quantum Science Technology (QSIT), European Unio Horizon 2020 research and innovation program within the Flagship on Quantum Technologies through GA 820445 (QIA) and under the Marie Sk\l{}odowska-Curie program through GA 675662 (QCALL).}

\bibliography{Manuscript.bib} 
\newpage

\appendix
\section{Coherence time automated evaluation} \label{App_A}

\begin{figure}
 \centering
  \includegraphics[width=0.75\columnwidth]{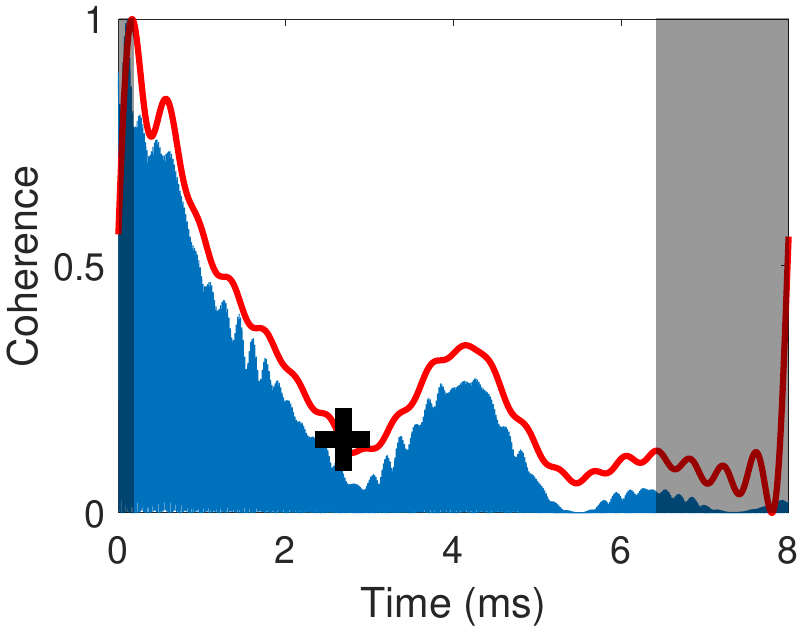}
  \caption{Comparison between the unfiltered coherence decay (in blue) in Pr:YSO for 0.7~mT, $\varphi=2.95$, $\theta=2.8$ and 3Y with the same echo after filtering out the Pr Zeeman frequency (in red). The gray areas around 0 and 7~ms are excluded from the decay time detection procedure. The black cross points out the end of the echo decay detected with the chosen threshold condition ($y=0.09$).}
  \label{fig:annx1}
\end{figure}

To infer a coherence time from our simulations, we automatically detect the end of the initial decay by choosing an arbitrary threshold condition for which we consider that the coherence is lost. As exposed in Sec.~\ref{Eu_sec} and Sec.~\ref{Pr_sec}, the non-Kramers ions coherence decay curves display highly contrasted modulations at the RE ion Zeeman frequency that cause the coherence to drop to zero periodically. However, such strong modulation cannot be seen as a loss of coherence due to the very strong revival of the coherence. With that consideration, using only a threshold condition on these curves is not sufficient to obtain a relevant characteristic decay time. We instead are interested on detecting the time at which the echo envelope goes below our threshold. Consequently, for non-Kramers ions we first apply a low-pass filter on the echo decay to get rid of the RE Zeeman frequencies if they are present. This is illustrated here in the case of Pr:YSO in Fig.~\ref{fig:annx1}.\par

Once the signal envelope is obtained, we discard the first 100~$\mu$s and last 15~\% of the signal, as the filtered signal in these time ranges frequently displays significant deviations from the unfiltered signal and causes errors on the detection (the grey areas on the figure \ref{fig:annx1} are the discarded parts of the echo). This is a direct consequence of the filtering function and unrelated to our simulations. To illustrate this time cutoff, for a 8~ms simulation the minimum decay time is 100~$\mu$s, and the maximum detected decay time is 6.5~ms.\par
Unfortunately, despite these precautions the simulation results may not resemble to an echo decay from which a coherence time could be deduced. For these curves, a coherence time of 0~ms is attributed by default (see e.g. zones with 0 coherence time in Fig.~\ref{subfig:71}). \par
However, for all valid decays, we choose a threshold condition for which we consider our coherence lost. The threshold value is an arbitrary choice based on visual evidence to have the most accurate decay time for the largest variety of cases. The chosen value for our threshold condition is an echo intensity going below 9~\%. \par

\begin{figure*}
\subfloat[\label{subfig:mean1}]{%
  \includegraphics[width=0.65\columnwidth]{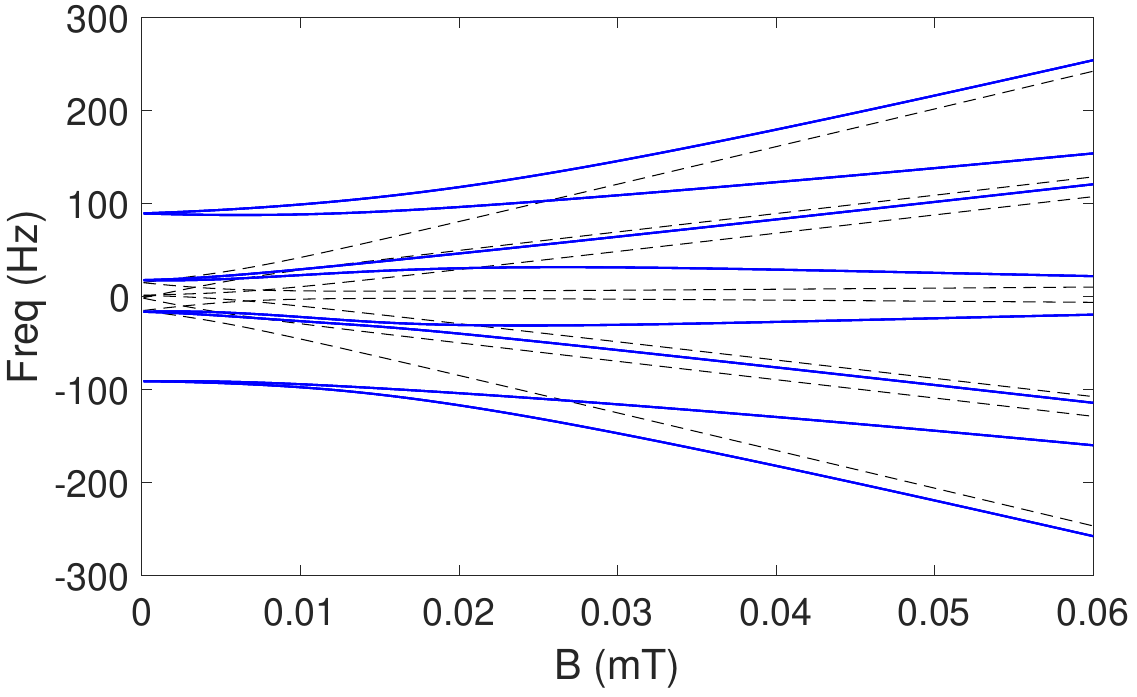}%
}\hfill
\subfloat[ \label{subfig:mean2}]{%
  \includegraphics[width=0.65\columnwidth]{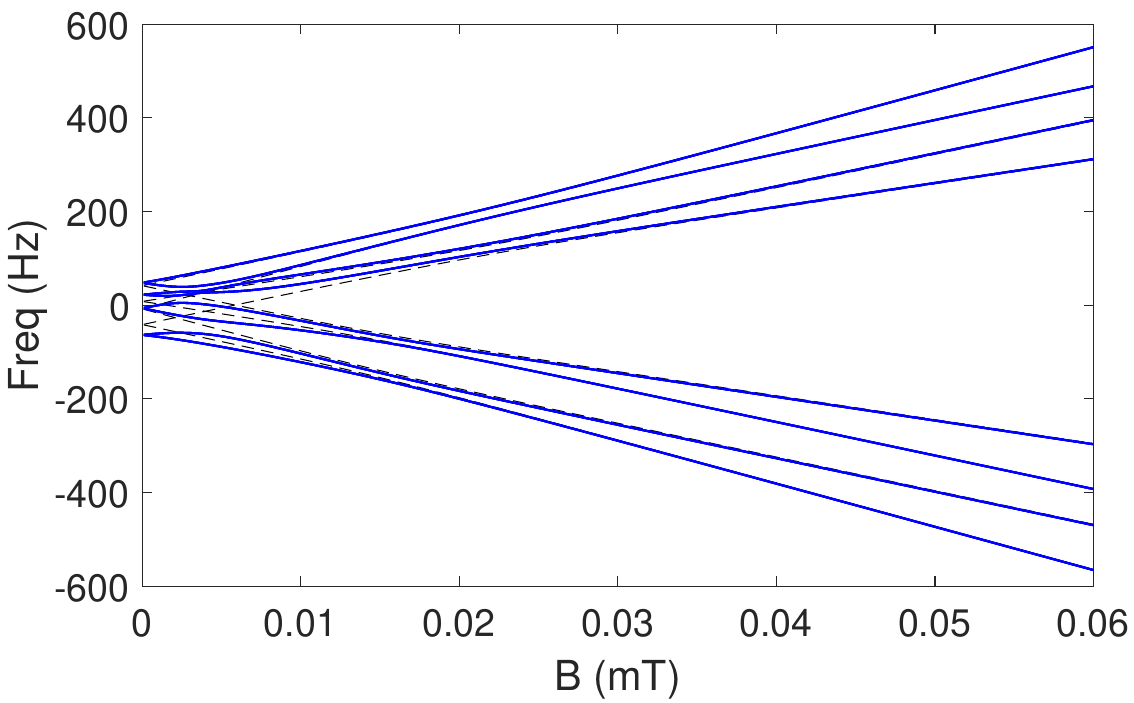}%
}\hfill
\subfloat[ \label{subfig:mean3}]{%
  \includegraphics[width=0.65\columnwidth]{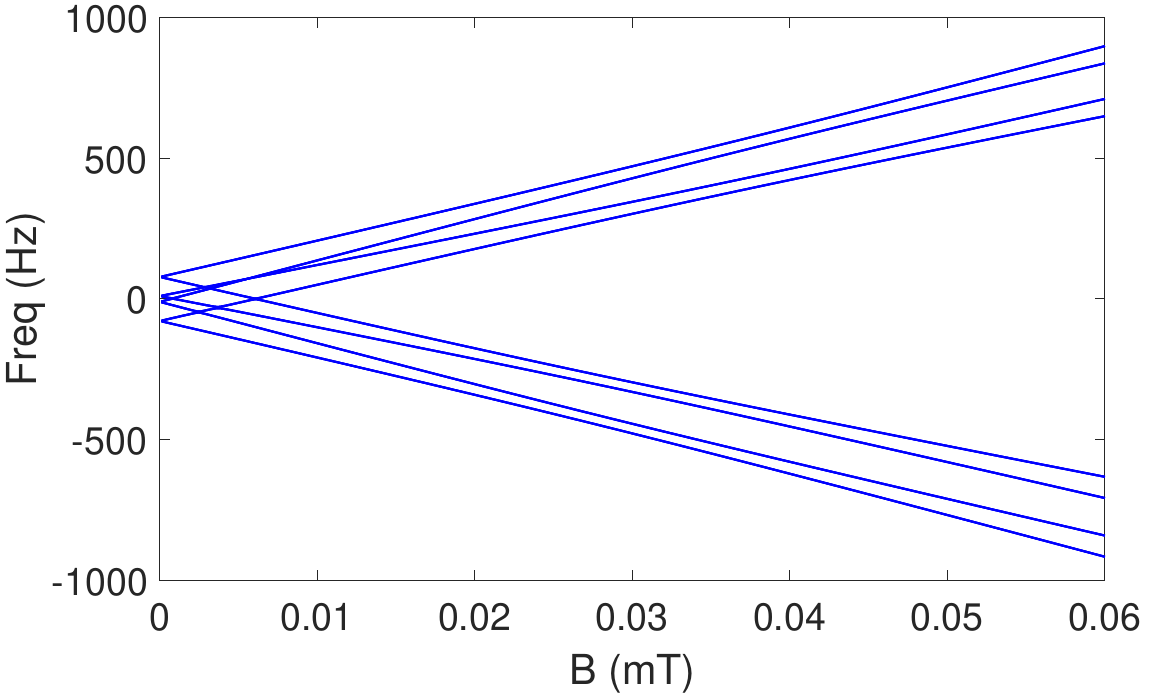}%
}\hfill
\caption{Eigenvalues $\lambda_l$ of the total Hamiltonian (\ref{eq:tot_hamilton}) as a function of the magnetic field amplitude, for one europium ion and the first two yttrium neighboring ions, for a field oriented along D$_1$, for a) $\ket{\pm1/2}_g$, b) $\ket{\pm3/2}_g$ and c) $\ket{\pm5/2}_g$. The two quickly separating branches stem from the Eu Zeeman contribution (separating $\ket{+k/2}$ from $\ket{-k/2}$), while the four-level manifold contain the two Y contributions. The mean values are subtracted for each hyperfine level for clarity.}
\label{fig:annx_dipole}
\end{figure*}
\section{Fermi Golden rule, mean dipole and CCE computations details} \label{App_B}
\label{Appendix:approxs}
\subsection{Fermi Golden rule}
\label{Appendix:Fermi}
Instead of computing exactly the coefficients in Eq.~(\ref{analyt_dyn}), the Fermi golden rule provides a faster and simpler method to derive them and express a decay rate. Several assumptions are needed in order to unfold the calculations. First, we assume that only one level is populated at the initial time. Secondly, the perturbation induced by the Hamiltonian $\hat{H}^{\rm dd}_{\rm RE-Y}$ must be weak compared to $\hat{H}_{0}$. The probability for $\hat{H}^{\rm dd}_{\rm RE-Y}$ to change the state of the system from the eigenstate $\ket{\varphi_{i}}_{g}$ to $\ket{\varphi_{f}}_{e}$ is:
\begin{equation}
\mathcal{P}_{if}= |\bra{\varphi_{f}}\hat{H}^{\rm dd}_{\rm RE-Y}\ket{\varphi_{i}}|^{2},
\end{equation}
which develops to
\begin{equation}\label{discrete_fermi}
\mathcal{P}_{if}= 4(\hat{H}_{\rm int})_{if}^{2}\left|\frac{\sin [\frac{(\omega_{f}-\omega_{i})t}{2\hbar}]}{(\omega_{f}-\omega_{i})}\right|^{2},
\end{equation}
with $\omega_{f}$ and $\omega_{i}$ the eigenfrequencies of the final and the initial state. \par
If the state $\ket{\varphi_{i}}$ is connected to $n$ levels, the total probability for the system to experience a state change is then:
\begin{equation}\label{sum_fermi}
\mathcal{P}=\sum_{k=1}^n \mathcal{P}_{ik}.
\end{equation}
If we now make the assumption that all the $f$ levels are spaced evenly, such that $\omega_{f}-\omega_{i}=k\delta$, and if we consider that the level $i$ is connected with a similar strength $(\hat{H}^{\rm dd}_{\rm RE-Y})_{if}=\nu$ to all $f$ levels, (\ref{discrete_fermi}) and (\ref{sum_fermi}) give:
\begin{equation}
\mathcal{P}= 4\nu^{2} \sum_{k} \frac{\sin^{2} [\frac{k\delta t}{2\hbar}]}{(k\delta)}^{2}.
\end{equation}
Given that:
\begin{equation}
\lim_{a\to 0} \sum_{k=-\infty}^{k=+\infty} \frac{\sin^{2} (a k)}{k}^{2}=\pi a,
\end{equation}
we obtain the following change of state probability per unit of time:
\begin{equation}\label{Fermi_rule}
\mathcal{P}= 2\pi \frac{\nu^{2}}{\delta}.
\end{equation}
This result is known as Fermi’s golden rule.\par
In our case, several of the assumptions needed to derive Eq.~(\ref{Fermi_rule}) do not hold. First, the coefficients of $\hat{H}^{dd}_{RE-Y}$ cannot be considered as constants for all pairs of levels. The RE-Y dipole-dipole interactions that generates them is strongly dependent on the Y ions distance and orientation, causing levels associated with a spin change of the closest Y ion to be strongly connected compared to farther ions. On a second hand, the energy levels cannot be considered as evenly spaced. This is directly illustrated with Fig.~\ref{spread}, that shows the Gaussian-like distribution of the Y manifold.

\subsection{Frozen central ion models}

\subsubsection{Model}
\label{annex:frozen}

In this paragraph, we detail the link existing between two models that can be found in the literature: the mean dipole moment model~\cite{Car2020,PhysRevB.100.165107} and the CCE model~\cite{Witzel06,PhysRevB.78.085315,PhysRevB.79.115320}. We also illustrate why they cannot be applied to our simulation cases.\\
Among others, the CCE method helps to estimate the time dependence of arbitrary echo sequences. In the case of a Hahn echo, it can be written of the form \cite{Witzel06}:
\begin{align}
\mathcal{E} (t)= \text{Tr}(\hat{\rho}_0e^{-i\hat{H}^{-}t}e^{-i\hat{H}^{+}t}e^{i\hat{H}^{-}t}e^{i\hat{H}^{+}t}),
\label{echo_tot}
\end{align}
where $\hat{H}^{-}$ (resp. $\hat{H}^{+}$) is the Hamiltonian of the bath (Y spins in our case) when the central spin is in its ground (resp. excited) state, and $\hat{\rho}_0$ is the initial spin bath state.\\
The dimension of the bath Hamiltonian makes the calculations quickly intractable if the number of ions is higher than a few. There comes the CCE into play, which consists in expressing the total echo as a product of contribution of individual clusters. In particular, at order 1, each cluster contains a single Y bath ion, and the total echo reads:
\begin{align}
\mathcal{E} (t)= \tilde{\mathcal{E}}_{\{1\}} (t)\tilde{\mathcal{E}}_{\{2\}} (t)\cdots\tilde{\mathcal{E}}_{\{N\}} (t),
\label{eq:echo_multip}
\end{align}
where 
\begin{align}
\tilde{\mathcal{E}}_{\{k\}} (t)&=\text{Tr}\left(\left(\hat{\rho}_k\right)_0e^{-i\hat{H}^{-}_kt}e^{-i\hat{H}_k^{+}t}e^{i\hat{H}_k^{-}t}e^{i\hat{H}_k^{+}t}\right)
\label{echo_k}
\end{align}
is the echo computed with the sole yttrium ion labeled $k$ in the bath and $\left(\hat{\rho}_k\right)_0$ its initial state. The Hamiltonians that shall be considered in expression (\ref{echo_k}) are then

\begin{align}
\hat{H}^{\pm}_k&=\hat{H}^{\rm Z}_{\rm Y_k}+\hat{H}^{\rm dd}_{\rm RE^\pm-Y_k}
\label{eq:two_Hamilts_CCE1}
\end{align}

Here, the bath Hamiltonian is solely given by $\hat{H}^{\rm Z}_{\rm Y_k}$, as at order 1 only single bath ions must be considered. The interaction Hamiltonian (\ref{eq:interaction_gal}) now reduces to the dipole-dipole coupling with yttrium ion labeled $k$, for the central ion in its ground ($\hat{H}^{\rm dd}_{\rm RE^--Y_k}$) or excited ($\hat{H}^{\rm dd}_{\rm RE^+-Y_k}$) state. These Hamiltonians are simply the restriction to the subspaces of central spin down or up of the magnetic dipole-dipole coupling, and are given by 

\begin{align}
\hat{H}^{\rm dd}_{\rm RE^\pm-Y_k}&=\bra{\pm}\hat{H}^{\rm dd}_{\rm RE-Y_k}\ket{\pm}=-\gamma_{\rm Y}\cdot\mathbf{I}\cdot\mathbf{B}_{\rm RE^\pm}^{\{k\}}
\end{align}
\begin{figure*}
\includegraphics[width=2\columnwidth]{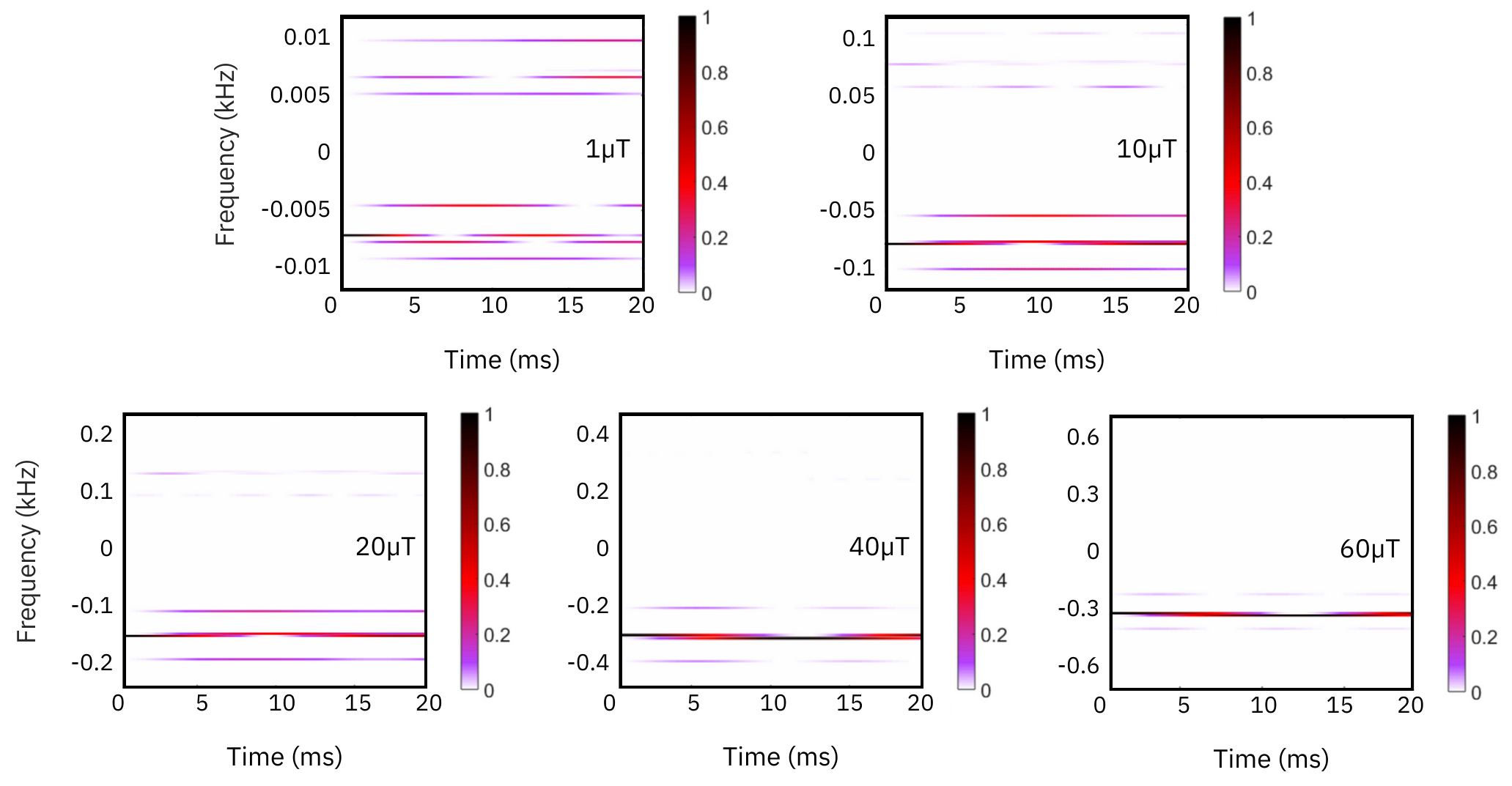}%
\caption{Population dynamics of the $\ket{\pm3/2}_g$ level of Eu:YSO as a function of the magnetic field amplitude (field along $D_1$).}
\label{fig:pop_dyn}
\end{figure*}
This turns to selecting only the diagonal blocks of the dipole-dipole interaction, in agreement with the hypothesis of a frozen central spin. Notice that this hypothesis is strictly equivalent to taking the mean value of the central spin dipole moment in the interaction Hamiltonian, allowing to define an equivalent magnetic field $\mathbf{B}_{\rm RE^\pm}^{\{k\}}$ radiated by the central ion at the position of the yttrium ion labeled $k$, as done in \cite{Car2018,Car2020}. With this in mind, we can re-write (\ref{eq:two_Hamilts_CCE1}) as:

\begin{align}
\hat{H}^{\pm}_{k}&=-\gamma_{\rm Y}\cdot\mathbf{I}\cdot\mathbf{B}_{\rm DC}-\gamma_{\rm Y}\cdot\mathbf{I}\cdot\mathbf{B}_{\rm RE^\pm}^{\{k\}}:=-\gamma_{\rm Y}\cdot\mathbf{I}\cdot\mathbf{B}^{\{k\}}_{\rm eff^\pm}.
\end{align}

By injecting this expression in (\ref{echo_k}) and taking initially perfectly mixed Y states $\left(\hat{\rho}_k\right)_0=1/2\ket{-}\bra{-}+1/2\ket{+}\bra{+}$, we find the expression
\begin{align}
\tilde{\mathcal{E}}_{\{k\}} (t)&=1-\frac{\sin^2(\theta_{k})}{2}\left[1-\cos\left(\Delta_k^-t\right)\right]\left[1-\cos\left(\Delta_k^+t\right)\right],
\label{eq:CCEtoMean}
\end{align}
with $\theta_{k}$ the angle between the two effective magnetic fields 
\begin{align}
\sin^2(\theta_{k})&=1-\mathbf{B}_{\rm eff^+}^{\{k\}}\cdot\mathbf{B}_{\rm eff^-}^{\{k\}}
\end{align}
and frequencies defined as
\begin{align}
\Delta_k^\pm&=\gamma_{\rm Y}||\mathbf{B}^{\{k\}}_{\rm eff^\pm}||.
\end{align}
Expression (\ref{eq:CCEtoMean}) coincides with the one that was used in~\cite{Car2020,PhysRevB.100.165107} in the mean field model.
Finally, the first order CCE of the echo is computed with (\ref{eq:echo_multip}).

\subsubsection{Invalidity of the approximation in the weak field regime}

In order to assess the validity of the frozen ion approximation, we plot the eigenvalues $\lambda_l$ of $\hat{H}$ (\ref{eq:tot_hamilton}) in the case of Eu:YSO for the three hyperfine ground states, as a function of the magnetic field amplitude, for a field orientated along the $D_1$ axis. Such eigenvalues are the ones dictating the dynamics of the evolution on the energetic ladder defined by $\hat{H}_0$. We remind here that the dynamical frequencies are given by the intervals $\Delta\lambda_{l,l'}=\lambda_l-\lambda_{l'}$. We also plot the same eigenvalues calculated with the mean dipole hypothesis in dashed black lines for assessing its validity. What can be directly witnessed is that in the magnetic field amplitude regime considered here, the frozen spin approximation is not always relevant, as strong energy mismatch (up to $\sim 100$~Hz for transitions of equivalent energy gap) are witnessed. Another important observation is the very different behavior of the hyperfine levels: while $\ket{\pm1/2}_g$ actively spin-flips with the neighboring Y ions, $\ket{\pm5/2}_g$ remains quite protected from this phenomenon, even at fields below 10~$\mu$T.

\section{Population dynamics in Eu}
\label{app:pop_field}
We show in this appendix the population dynamics of level $\ket{\pm3/2}_g$ in Eu:YSO as a function of the magnetic field amplitude, to highlight spin flip processes at stake at different field amplitudes. To this extent we consider a Eu ion in interaction with the first two neighboring Y ions, and place the system initially in an eigenstate of $\hat{H}_0$. Therefore, on the contrary to the sequence used to simulate coherence evolution over time in the body of the paper, no RF pulse is applied to create superpositions between the hyperfine states. We then let the system to evolve under dipole-dipole interaction only and plot the population of each level versus time. This is shown in Fig.~\ref{fig:pop_dyn} for a field applied along $D_1$, with a magnitude ranging from 1 to 60~$\mu$T. The population is shown with a colorscale, and reveals that at fields lower than 10~$\mu$T essentially all states end up being populated over time. Above 10~$\mu$T, we see that the Eu spin flip begins to be forbidden, and mostly Y spin flips and Y flip-flops remain. Then, above 40~$\mu$T, only two levels are populated, corresponding to Y-Y flip flops. This observation corroborates the observations done in coherence with simple population simulations.

\section{Coherence map of Yb}
\label{app:Ybcoh}
\begin{figure*}
\centering
\subfloat[\label{fig:3Y_Yb_2D_map}]{%
  \includegraphics[width=0.7\columnwidth]{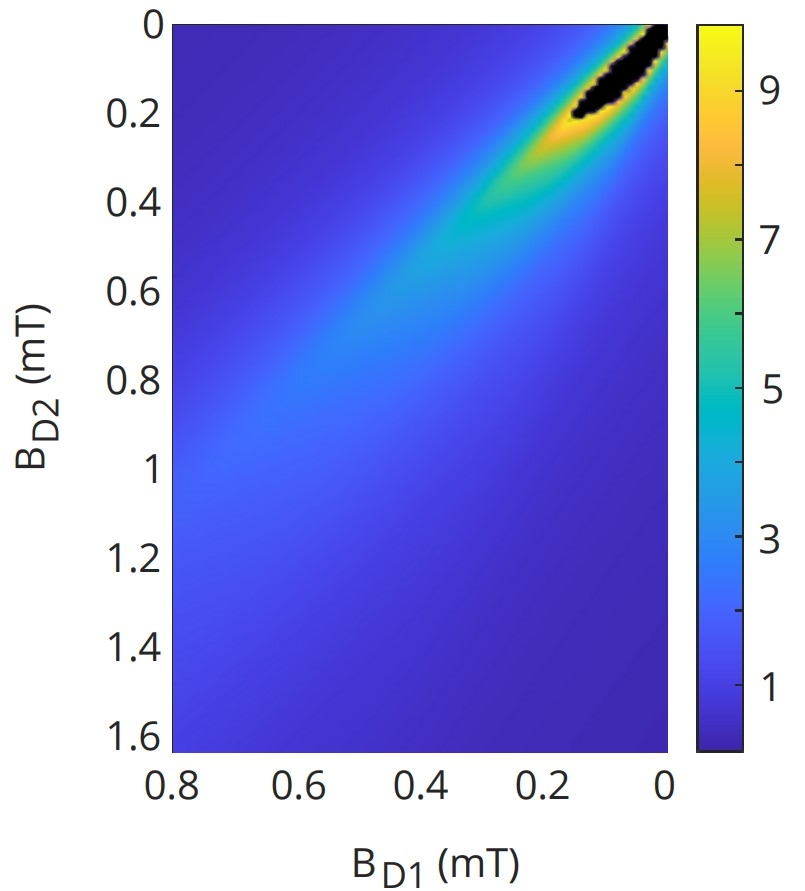}%
}\hfill
\subfloat[\label{fig:Hahn}]{%
  \includegraphics[width=1\columnwidth]{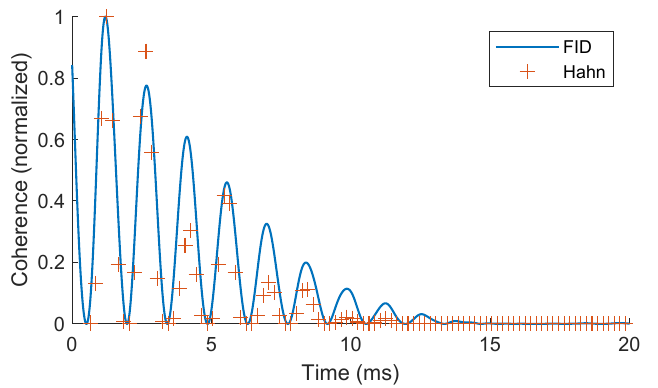}%
}\hfill
\caption{a) Coherence time in millisecond as a function of magnetic field angle, for one Yb ion and 3 Y ions. b) Decays simulated with the simple FID sequence (used in the paper, solid blue), and with a Hahn sequence (orange dots), for 3 Y ions.}
\label{fig:appfin}
\end{figure*}
In order to get a general picture of the coherence dependence of Yb:YSO, we plot in Fig.~\ref{fig:3Y_Yb_2D_map} the echo decay times (see App.~
\ref{App_A} for its estimation), as a function of the magnetic field direction and amplitude for one Yb ion and the first three neighboring Y ions. \par
The magnetic dipole of Yb scales linearly with the magnetic field amplitude, therefore Yb has a zero-field ZEFOZ point, where the coupling with the surrounding Y ions is minimal~\cite{ortu_simultaneous_2018}. From that point, the coherence time gets longer than the simulated evolution time ($10~$ms), causing the automated decay time detection to fail (black region on the plot). Past that zone, the simulated decay time decreases at higher magnetic field amplitudes because of the increased Yb-Y couplings. We also notice that the coherence time is significantly improved along a specific direction. It corresponds to the direction for which the first order transition gradient $||\mathbf{S}_{1}||$ is minimal in the $(D_{1}, D_{2})$ plane, at $\phi=55.9^{o}$. This has also been witnessed experimentally \cite{Nicolas2022}.

\section{Relevance of the sequence considered}
\label{app:Hahn}

Here, we compare the simulated decay for europium at 60 $\mu$T (case of Fig.~\ref{fig:Eu_echoes}) with a decay generated for a full simulation of a Hahn echo sequence. Both simulations are shown in Fig.~\ref{fig:Hahn}, and even if slight deviations from one curve to the other can be seen, the overall shape, modulation pattern and characteristic decay time are similar. This curve comforts us in the relevance of our approach, even at such field, for which single ion inhomogeneous decoherence is close to be limiting our decay time.\par

\end{document}